\documentclass{egpubl}
\usepackage{egsr2020}

\SpecialIssuePaper

\usepackage[T1]{fontenc}
\usepackage{dfadobe}  

\usepackage{cite}  
\BibtexOrBiblatex
\electronicVersion
\PrintedOrElectronic
\ifpdf \usepackage[pdftex]{graphicx} \pdfcompresslevel=9
\else \usepackage[dvips]{graphicx} \fi

\usepackage{egweblnk} 

\usepackage[utf8]{inputenc}
\usepackage[english]{babel}
\usepackage{amsmath}
\usepackage{amsfonts}
\usepackage{placeins}
\usepackage{amssymb}
\usepackage{hyperref}
\usepackage{graphbox}
\usepackage[normalem]{ulem}
\usepackage{rotating}
\usepackage{multirow}
\usepackage{overpic}
\usepackage[dvipsnames]{xcolor}

\definecolor{todoCol}{rgb}{1,0.5,0}
\definecolor{GDCol}{rgb}{0,0.9,0.3}
\newcommand{\NEW}  [1] {{#1}}

\title{Guided Fine-Tuning for Large-Scale Material Transfer}

\author[V. Deschaintre, G. Drettakis \& A. Bousseau]
{\parbox{\textwidth}{\centering Valentin Deschaintre$^{1,2,3}$, George Drettakis$^{1}$ and Adrien Bousseau$^{1}$ \\ \vspace{4mm}
$^1$ Université Côté d'Azur, Inria  $^2$ Imperial College London \\$^3$ Optis for Ansys\\}
}

\begin{document}
\teaser{
\begin{overpic}[scale=1.]{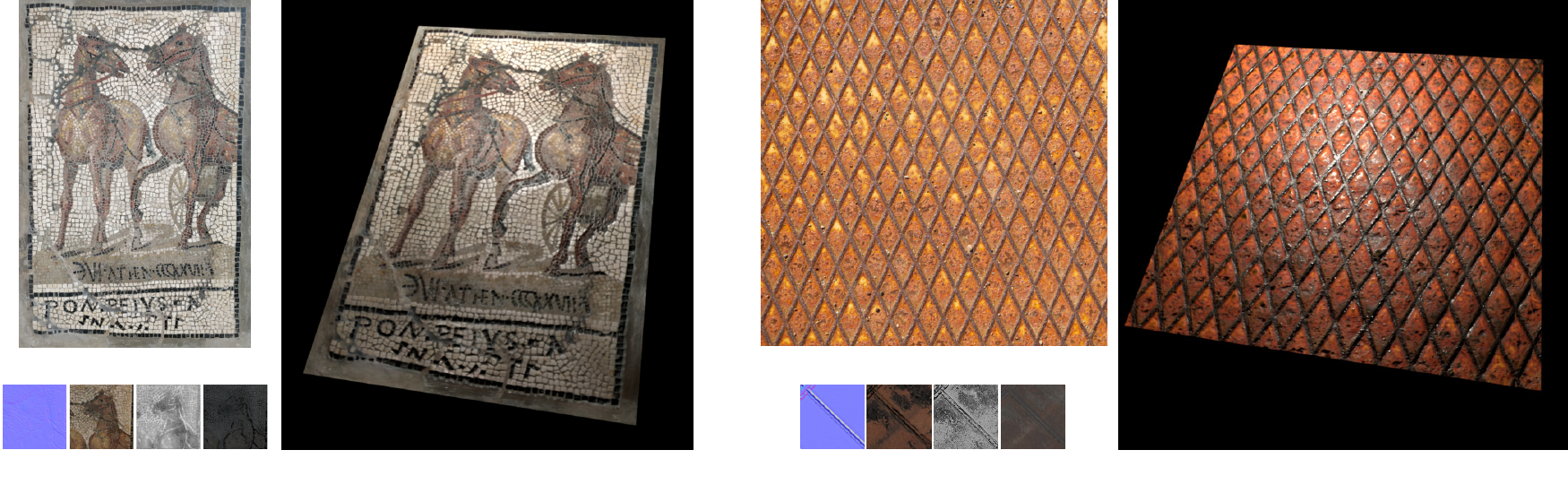}
\put(5.5,8.4){\small{HD Input}}
\put(3,1.5){\small{SVBRDF exemplar}}
\put(28,1.5){\small{Rendering}}
\put(57,8.4){\small{HD Input}}
\put(53.5,1.5){\small{SVBRDF exemplar}}
\put(82.5,1.5){\small{Rendering}}
\end{overpic}
\vspace{-2mm}
	\caption{Our method transfers the appearance of one or a few exemplar SVBRDFs to a target picture. This approach allows the capture of large planar surfaces taken with ambient lighting (far left), by extracting the SVBRDF exemplars from close-up flash pictures (lower left), as well as the creation of plausible SVBRDFs from internet pictures by using existing artist-designed materials as exemplars (right). Please see supplemental materials for high-resolution SVBRDF parameter maps and animated renderings of all our results, which give a much better impression of the material properties. } 
\label{fig:teaser}
}

\maketitle
\begin{abstract}
We present a method to transfer the appearance of one or a few exemplar SVBRDFs to a target image representing similar materials. Our solution 
is extremely simple: we fine-tune a deep appearance-capture network on the provided exemplars, such that it learns to extract similar SVBRDF values from the target image. 
We introduce two novel material capture and design workflows that demonstrate the strength of this simple approach.
Our first workflow allows to produce plausible SVBRDFs of large-scale objects from only a few pictures. Specifically, users only need take a single picture of a large surface and a few close-up flash pictures of some of its details. We use existing methods to extract SVBRDF parameters from the close-ups, and our method to transfer these parameters to the entire surface, enabling the lightweight capture of surfaces several meters wide such as murals, floors and furniture. In our second workflow, we provide a powerful way for users to create large SVBRDFs from internet pictures by transferring the appearance of existing, pre-designed SVBRDFs. By selecting different exemplars, users can control the materials assigned to the target image, greatly enhancing the creative possibilities offered by deep appearance capture. \textcolor{red}{\textbf{This is a very low resolution version of the paper, you can find a more reasonably compressed version on the project page: \url{https://team.inria.fr/graphdeco/projects/large-scale-materials/}}}
\end{abstract}




\section{Introduction}  
Recent progress on lightweight appearance capture allows the recovery of plausible real-world spatially-varying reflectance distribution functions (SVBRDF) from just a few photographs of a surface. In particular, multiple methods take as input one or several photographs captured with a hand-held camera, where the co-located flash provides informative spatially-varying illumination over the measured surface sample \cite{Aittala15,Aittala16,Riviere2016,Hui2017,Deschaintre18,li18,Deschaintre19,gao19}. However, near-field flash lighting greatly restricts the \emph{scale} at which materials can be captured -- typically a dozen centimeters wide using a cell phone held at a similar distance. Relying on a flash also prevents these methods from processing existing images captured under \emph{unknown lighting}, such as textures downloaded from the Internet. Finally, another common limitation of the above methods is that they rely on black-box optimization or deep learning to infer SVBRDF parameters from few measurements, offering little \emph{user control} on their output. 
We address all three limitations by proposing a \emph{by-example} appearance capture method, which recovers SVBRDF parameter maps over large surfaces captured under environment lighting by transferring information from one or a few \emph{exemplar SVBRDF patches} (Fig.~\ref{fig:teaser}), that can either be extracted from additional close-up flash photos, or come from a database of SVBRDFs.

Our technical solution to transfer material appearance from exemplars is surprisingly simple yet extremely effective. We build on a state-of-the-art SVBRDF capture deep network \cite{Deschaintre18}, which 
we re-train to take as input a single image captured under environment lighting, and output SVBRDF maps (normals, diffuse albedo, specular albedo, and roughness). 
Our key idea is to fine-tune this network on the provided exemplars, which strongly biases the network towards their specific SVBRDF values \NEW{using the available color and texture cues}. We then run this custom network on the target image, which effectively produces SVBRDF maps that contain similar values to the ones of the exemplars.

However, naively fine-tuning a large deep network on a small number of exemplars results in dramatic over-fitting, as the network quickly memorizes the spatial layout of the exemplars rather than learn material-specific filters that would generalize to the target image. We address this challenge by carefully augmenting the exemplar set. In particular, we generate a unique training image for each iteration of the fine-tuning by applying random geometric transformations on the exemplars, and by combining multiple transformed exemplars into a single collage via random masks. Our experiments demonstrate that this augmentation is critical to the success of the method.

We introduce two new \NEW{applications} that demonstrate the strength of our approach. 
Our \emph{on-site acquisition} scenario is the first \NEW{application} to allow capture of plausible material properties of \emph{large} surfaces with just a few photos. In this case, we capture a single photograph of a large surface as well as one or a few close-up flash photographs of its details. We then use an off-the-shelf network to extract SVBRDF maps from the flash photographs, and use our fine-tuned network to transfer this information to the large image, effectively acquiring SVBRDFs several meters wide. In our second scenario -- \emph{creative design} -- we provide a powerful method for users to create realistic SVBRDFs from stock photos,
simply using artist-created SVBRDFs downloaded from the Internet as exemplars.
This demonstrates how our method allows fine control on the design process for SVBRDFs.







In summary, this paper makes the following contributions:
\begin{itemize}
\item We present a simple yet very effective algorithm to transfer material appearance from a few exemplars to a target image.
\item We introduce a lightweight method to capture SVBRDFs of large planar surfaces, based on this algorithm. 
\item We introduce a novel workflow that allows material designers to create new SVBRDFs from existing photos and SVBRDF patches (\emph{e.g.}, taken from online texture and SVBRDF repositories), 
using the same algorithm. 

Our code, data and supplemental material are available here: \url{https://team.inria.fr/graphdeco/projects/large-scale-materials/}
\end{itemize}


\begin{figure*}
\includegraphics[width=\linewidth]{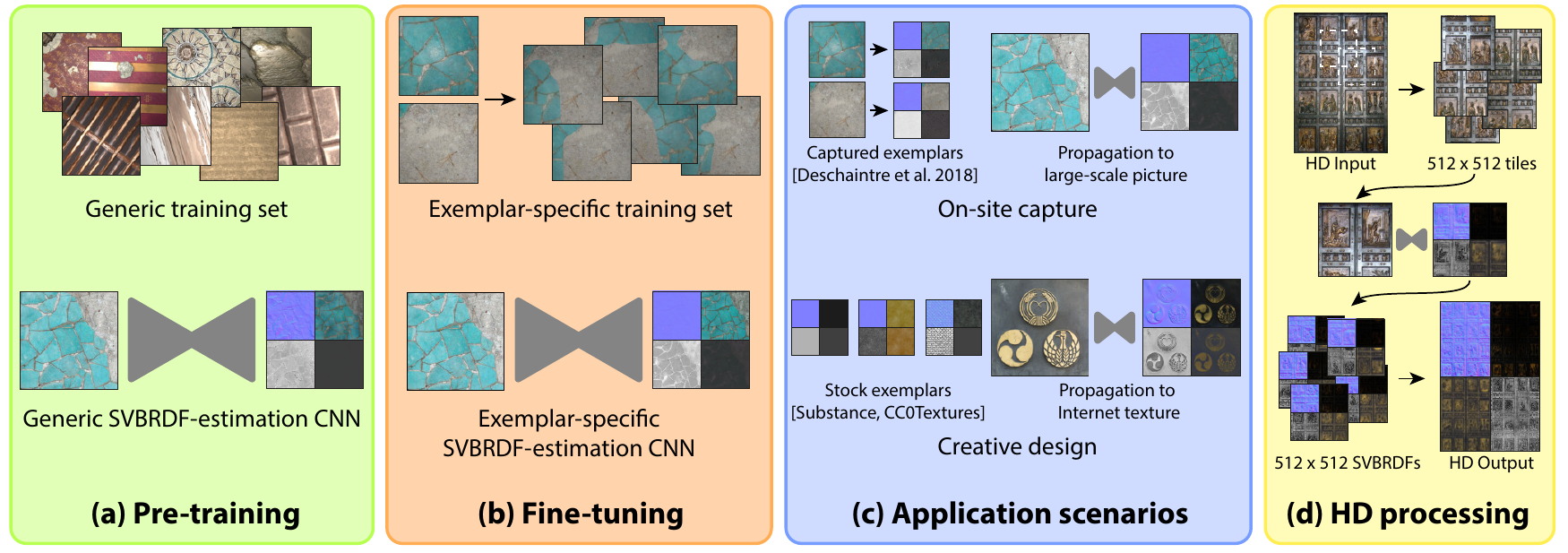}
\vspace{-3mm}
\caption{Main steps of our method. We first pre-train an SVBRDF prediction network \cite{Deschaintre18} on a large set of synthetic SVBRDF maps rendered under varying lighting (a). While this generic network produces plausible results, it often mis-interprets the material features in the absence of flash cues. Our key idea is to fine-tune the pre-trained network on renderings of user-provided SVBRDF exemplars (b). After fine-tuning, the resulting network combines generic pre-training knowledge with information from the exemplars. Here, this allows our method to interpret the cyan tiles as more shiny than the grey concrete. We demonstrate this approach on two application scenarios, either to acquire large-scale real-world surfaces by propagating small-scale exemplars (c, top), or to design new SVBRDFs by propagating existing SVBRDF maps over internet textures (c, bottom). While we train our network on images of $512\times 512$ pixels, we process HD images of more than $2048 \times 2048$ pixels by processing small tiles individually, and by stitching their predicted SVBRDFs to generate the final output. This is made possible by the absence of strong local flash highlights in the input image.}
\label{fig:overview}
\end{figure*}

\section{Related Work}
Appearance capture and design is a vast and active research field; We refer to the survey by Guarnera et al~\shortcite{guarnera16} for a general introduction, and to the one by Dong~\shortcite{dong19} for a focus on methods based on deep learning. Here we discuss lightweight SVBRDF capture methods most similar to our approach, as well as related work on by-example image synthesis and deep learning.

Reconstructing multiple SVBRDF maps from one or a few pictures is an ill-posed problem, as the radiance observed in the pictures can be explained by a number of different combinations of SVBRDF parameters. Existing work tackled this challenge by incorporating domain-specific priors on the solution, either designed by hand or learned from large quantities of SVBRDF data. Example hand-designed priors include the assumption that the material sample is stochastic or self-repetitive \cite{Wang2011,Aittala15,Aittala16}, or that the lighting exhibits natural statistics \cite{Dong14} and physical properties \cite{Riviere17}. Data-driven methods seek to explain the observed data as a combination of known BRDFs \cite{Hui2017,Ren11}, or more recently by training deep neural networks to predict SVBRDF parameter maps using synthetic data for supervision \cite{Deschaintre18,li18,Deschaintre19,gao19}. While the above methods target planar surfaces like ours, some have also been extended to the problem of jointly capturing shape and material appearance, either using inverse-rendering optimization \cite{Baek18,Nam18} or deep learning \cite{LiShapeRefl:2018}. 

Most of these methods succeed in the task by targeting flash pictures captured at a small distance from planar material samples or small curved objects. 
In such a configuration, the flash produces a highlight at the center of the image as well as diffuse shading on its boundary, which provides information about the specular and diffuse behavior of the surface respectively, as well as complementary cues about normal variations. However, the use of a flash imposes three limitations for such methods. 
First, capturing large-scale surfaces would require the use of a large, powerful flash, defeating the purpose of these lightweight methods. 
Second, because flash lighting yields different visual cues in different places of the image, existing methods need to process the image in its entirety to aggregate all information, which is problematic for deep learning methods as the network resolution is limited by the GPU memory -- related methods were typically demonstrated on images of $256 \times 256$ pixel resolution. \NEW{In supplemental material we provide an example showing how the method by Deschaintre et al. \shortcite{Deschaintre18}, trained at low resolution, degrades when applied at higher resolution since the relative network footprint is reduced compared to the size and location of the flash highlights.}
Third, the reliance on co-located flash lighting prevents these methods from handling images taken in the wild with unknown lighting and arbitrary scale.
Our approach lifts all these limitations thanks to SVBRDF exemplars that bias the interpretation of the image towards specific material values, effectively alleviating the need for the visual cues offered by flash lighting. 

In contrast to the above methods, Li et al. \cite{Li17} proposed a deep network capable of predicting SVBRDFs fom images captured under environment lighting, including images taken in the wild. However, environment lighting alone provides little in terms of visual cues of the complex behavior of SVBRDFs, which makes their results inferior to the ones obtained by more recent flash-based methods. In particular, their method assumes that the specular term of the BRDF does not vary spatially, while spatially-varying roughness greatly contributes to the richness of real-world materials.
Li et al. \cite{Li17} also introduced the concept of \emph{self-augmentation}, which was further studied by Ye et al. \cite{Ye2018}. The idea is to use the network output to build new training samples, effectively augmenting the \emph{diversity} of SVBRDFs seen by the network. 
This strategy differs from ours, since our goal is rather to \emph{specialize} the network to extract user-provided SVBRDF values, which we achieve by fine-tuning the network on specific exemplars.

Our use of exemplar images makes our problem akin to \emph{image analogies} \cite{Hertzmann2001Image}, where the goal is to copy the appearance of an exemplar onto a target. The image analogies framework has been applied to a variety of problems, such as image colorization \cite{Welsh2002}, style transfer \cite{Fiser16-SIG}, texture transfer \cite{TexSynth2015}. All these methods share the strength of providing high-level control on their output thanks to the exemplars, a feature that we now provide in the context of SVBRDF capture and design. Closer to our application domain is the work by Melendez et al.~\shortcite{Melendez12}, who used patch-based texture synthesis to transfer diffuse albedo and depth variations from small material exemplars to large fa\c{c}ade images. However, their approach assumes that every pixel of the target can be put in correspondence to similar pixels of the exemplar, which yields visual artifacts when the exemplars do not contain all the material variations of the target image (see Fig.~\ref{fig:comparisonAdaINPatchMatch}). Several recent \NEW{methods use deep learning for image-to-image translation problems in supervised \cite{isola17, wang18} or unsupervised settings \cite{zhu17}. In particular, multiple} methods combine dense correspondences with deep learning to achieve more robust colorization \cite{He2018, He2019} and style transfer \cite{Liao2017}.
Our solution is simpler as it does not require explicit correspondences between the exemplars and the target. Instead, we train a deep material capture network to learn the mapping between the colors and textures of the exemplars and their SVBRDFs values, allowing us to apply this mapping on the target. \NEW{In concurrent work, Texler et al.~\shortcite{texler20} used a similar strategy to specialize a style transfer network using a small number of style exemplars.}

By complementing an input image with a few user-provided exemplars, our approach also relates to the interactive material design system \emph{AppGen} \cite{dong2011appgen}. The main difference between the two approaches resides in the level of expertise required and control offered. While AppGen offers fine control on the local interpretation of an image thanks to user scribbles, it requires users to manually segment the different materials in the image, and to specify each specular BRDF. 
In contrast, users of our approach need only select exemplar SVBRDFs from an existing library, or acquire them using an existing lightweight method, and let our method automatically transfer BRDF values from the exemplars to the target image. Our on-site acquisition scenario also follows the same two-scale capture strategy as \emph{Manifold Bootstrapping} \cite{Dong2010}, although we only need a few pictures of the surface at small and large scale where Dong et al. rely on specialized hardware to capture local BRDF samples, and on multiple photographs under varying lighting to capture global appearance.

Our technical solution for material transfer is inspired by the recent concept of \emph{internal learning}, 
i.e., training a deep neural network on a specific image rather than on a large dataset. This intriguing idea first appeared in the seminal work of Ulyanov et al. \shortcite{Ulyanov_2018_CVPR} on \emph{deep image priors}, where a network trained to reconstruct a specific image was shown to denoise or inpaint that image. Subsequent work used image-specific training for various tasks, including unsupervised super-resolution \cite{ZSSR} and GAN-based image editing \cite{Bau:Ganpaint:2019,Shaham_2019_ICCV}. Our approach differs, since while we fine-tune a deep network on a small set of images, we use the resulting network to \emph{transfer} the knowledge it acquired on a different target image.
Our work also relates to the \emph{TileGAN} method of Fr\"{u}hst\"{u}ck et al.~\cite{Fruehstueck2019TileGAN}, who train a conditional GAN to perform small-scale texture synthesis, and apply this GAN in a sliding-window fashion to produce large-scale images. However, training a GAN to synthesize a specific texture takes several days, while we show that it takes only a few minutes to fine-tune a generic material acquisition network to achieve successful material transfer.
Our strategy can also be seen as a form of \emph{few-shot learning}, that aims at adapting a pre-trained model to a new category of data given only a few examples of such data \cite{Liu_2019_ICCV}. As mentioned above, in our context, a few minutes of fine-tuning on augmented exemplars is sufficient to achieve this adaptation.




\section{Method}
Fig.~\ref{fig:overview} provides a visual overview of our method to extract SVBRDF parameter maps for large-scale surfaces. The main steps include pre-training a deep SVBRDF prediction network on a varied set of SVBRDFs (Fig.~\ref{fig:overview}a), fine-tuning this network on our exemplars (Fig.~\ref{fig:overview}b), and finally using this exemplar-specific network to extract SVBRDFs similar to the exemplars over images of large surfaces, either captured on site or downloaded from the Internet (Fig.~\ref{fig:overview}c). We first describe typical inputs to our method, before explaining how we pre-train and fine-tune the deep network to achieve material transfer.


\subsection{Inputs}

Our goal is to generate SVBRDF parameter maps for large-scale planar surfaces, such as walls, doors or furniture.
To do so, our method takes two forms of input. First, a single picture of the surface of interest, captured under ambient indoor or outdoor lighting. Second, a series of SVBRDF patches that represent small parts of the surface, or of a similar material. To obtain these patches, we either capture close-up flash pictures of the surface and run an existing single-image SVBRDF method \cite{Deschaintre18}, or we select SVBRDFs from libraries of artist-designed materials \cite{Substance,CC0Textures} (Fig.~\ref{fig:overview}c).

As a pre-process, we split the large-scale image into tiles of $512 \times 512$ pixels. Our method processes each tile independently, and generates the final output by stitching these individual predictions (Fig.~\ref{fig:overview}d, Sec.~\ref{sec:stitching}). Neighboring tiles have an overlap of $256$ pixels to facilitate subsequent stitching of their SVBRDF maps. Applying the network in this sliding-window fashion ensures that our method has a constant memory footprint, and as such scales to images of arbitrary resolution. In contrast, while running the network in a fully-convolutional manner would also allow the processing of images of varied resolution \cite{gao19}, the memory consumption of the method would increase with resolution, and eventually saturate GPU memory.

Note also that we assume that all tiles receive approximately the same lighting, which is not the case for pictures taken with a flash as used in prior work \cite{Deschaintre18,li18,Deschaintre19,gao19}.


\subsection{Neural network pre-training}
Our method processes each tile of the input image independently to output four Cook-Torrance SVBRDF maps \cite{Cook82}, corresponding to the normal, diffuse albedo, specular albedo, and specular roughness of each input pixel. We perform this task with the convolutional neural network proposed by Deschaintre et al. \cite{Deschaintre18}. While the original network was trained with synthetic images rendered under flash lighting, we re-train it with images rendered under a random directional light to be robust to arbitrary lighting conditions in our inputs. We also mimic a simple white sky dome by adding a small multiple of the diffuse and specular albedos to the renderings, which we found to be necessary to prevent metallic materials to appear completely dark away from the specular highlight. Despite its simplicity, we found this lighting model to work well on real-world pictures, including textures downloaded from the internet (Sec.~\ref{sec:eval}).
We generated our training data with the same set of parametric SVBRDFs as Deschaintre et al., except that we render them at a higher resolution to train the network to process images of $512 \times 512$ pixels. In total, the network is pre-trained for 800.000 iterations, which took around 8 days on a 1080TI graphics card.

Pre-training the network on a large set of SVBRDFs not only accelerates the subsequent fine-tuning step, it also equips the network with general priors on material appearance, which complements the exemplar-specific priors learned during fine-tuning (see Fig.~\ref{fig:onlineResults}).

\subsection{Neural network fine-tuning}
\begin{figure}
\includegraphics[width=.92\linewidth]{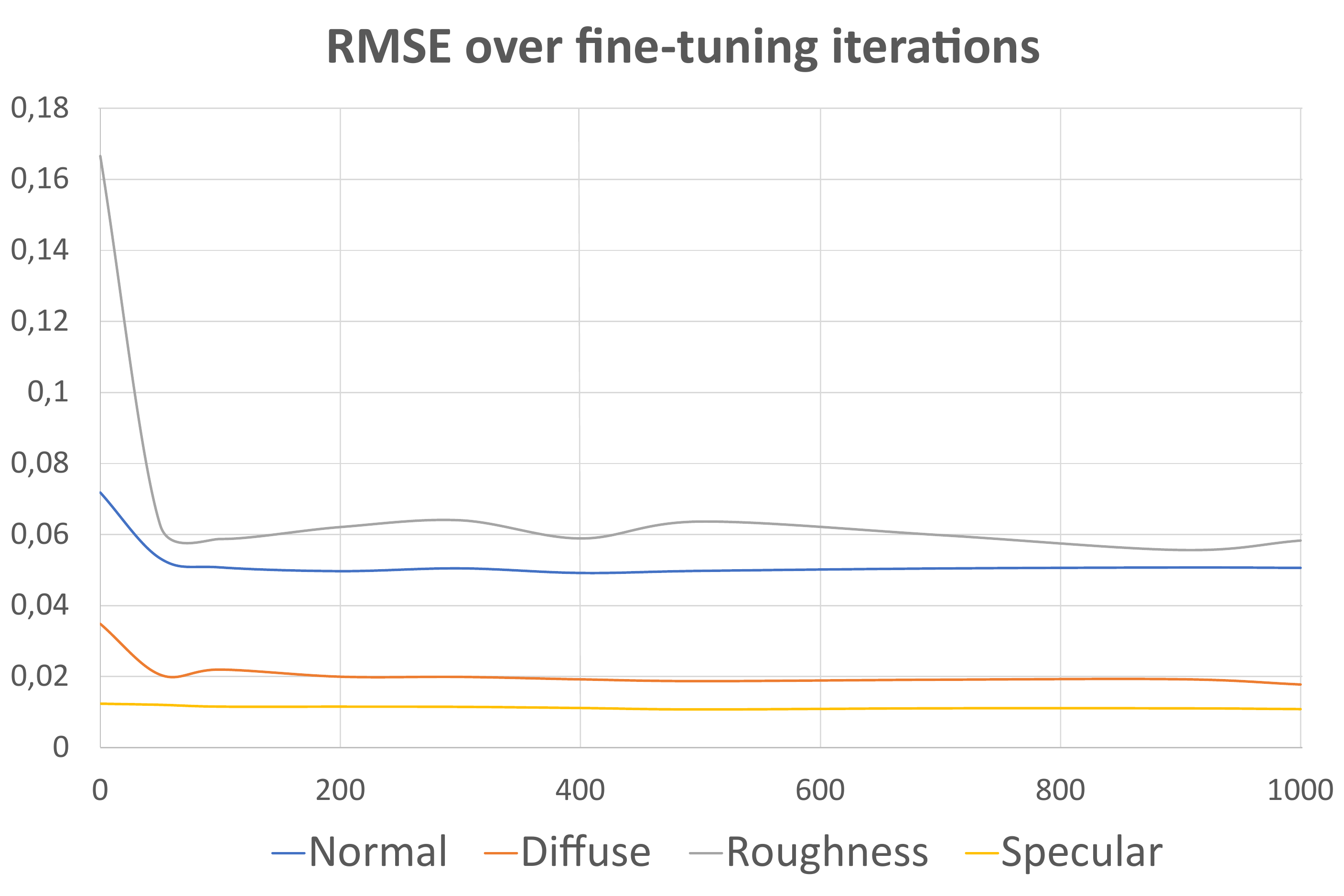}
\vspace{-3mm}
\caption{Average RMSE of predicted maps for $4$ synthetic SVBRDFs, using crops of these SVBRDFs as exemplars. The error quickly drops in less than $100$ training iterations.}
\label{fig:plot_convergence}
\end{figure}

A single image often does not provide enough information to recover SVBRDF parameters unambiguously, especially in the absence of flash highlights. The key idea of our work is to favor the SVBRDF parameter values present in the exemplars by fine-tuning the network on these images. In other words, we perform a number of training iterations where we ask the network to predict exemplar SVBRDF maps given a rendering of that SVBRDF as input. The network thus becomes increasingly specialized in mapping the color and texture of the exemplar renderings to their normal and reflectance values. 
We used $1000$ training iterations for all our results, which takes around $2$ minutes on a 1080 Ti GPU and is largely sufficient to achieve successful transfer. Our numerical experiments suggest that most of the improvement occurs within a few hundred iterations (Fig.~\ref{fig:plot_convergence}). Once fine-tuned, we run the network on each input tile to obtain its SVBRDF maps.

\subsection{Exemplar augmentation}
\begin{figure}
\begin{overpic}[scale=0.23]{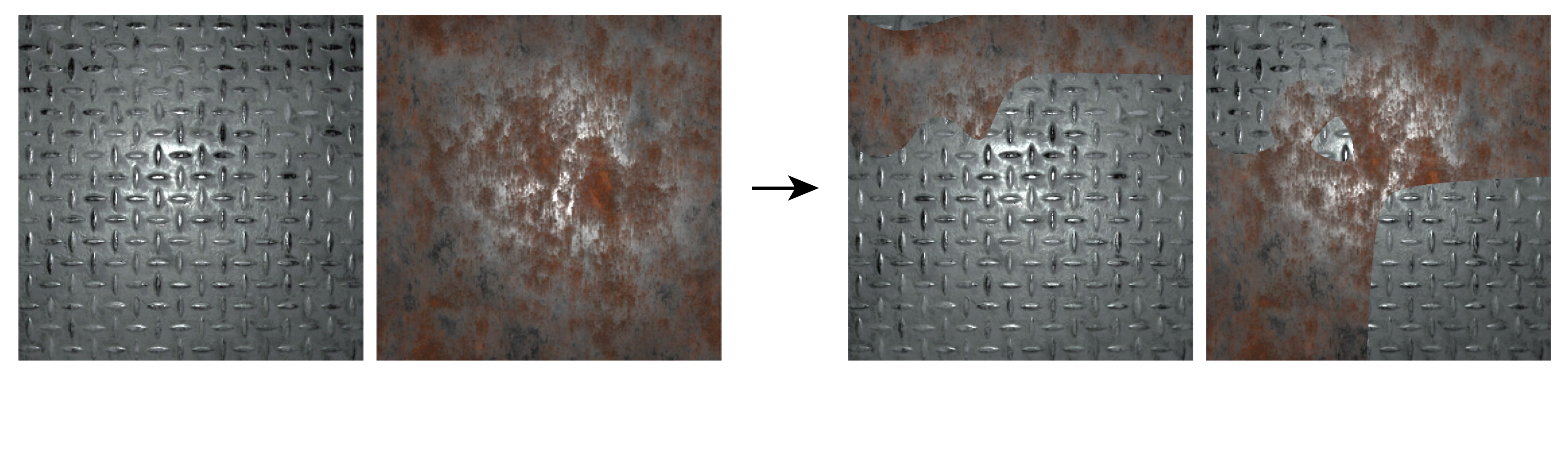}
\put(11,2){Input exemplars}
\put(59,2){Augmented exemplars}
\end{overpic}
\vspace{-2mm}
\caption{We augment the input set of SVBRDF exemplars by composing them using a low-frequency random mask.}
\label{fig:mixed_exemplars}
\end{figure}

While extremely simple, the above procedure quickly overfits the network so much on the few exemplars that it does not generalize to input images having a different distribution of materials regions. Our solution to this challenge is to apply massive data augmentation on the exemplars to obtain a training set that retains their local appearance, but varies their overall layout. We achieve this goal by generating, for every training iteration, a unique SVBRDF that is composed of pieces of two randomly-selected different exemplars. We first apply random scaling and cropping on these exemplars, and then combine them according to a binary mask that we generate by thresholding a low-frequency Perlin noise (Fig.~\ref{fig:mixed_exemplars}). We perform all these processing steps at training time in TensorFlow \cite{tensorflow2015} to reduce storage and data transfer. When only one exemplar is provided, we only augment it with scaling and cropping. We use the same lighting model as for pre-training to render this training set.

\subsection{Post-processing}
\label{sec:stitching}
The last step of our method consists in merging the predictions of all tiles into a large-scale SVBRDF. Since all tiles are processed using the same exemplars, neighboring tiles mostly agree in their predictions up to low frequency variations. We achieve a seamless composite by blending the tiles over their overlap using a Gaussian weighting kernel that gives a weight of $1$ at the center of the tile and reaches almost $0$ at its border. This mechanism allows our method to be applied on \emph{high-resolution inputs of arbitrary aspect ratio}, as shown in our results of up to $2048 \times 2048$ pixels.

\begin{figure*}
\begin{tabular} {cccc}
HD input picture & Examplars & Results & Rendering \vspace{1mm} \\
\hspace{-4.0mm} \includegraphics[align=c, width=0.32\linewidth]{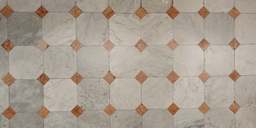} & \begin{tabular} {cc}
\hspace{-4mm} \begin{sideways} \hspace{-3mm} \tiny{Norm} \end{sideways} & \hspace{-4.0mm} \includegraphics[align=c, width=0.035\linewidth]{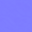} \vspace{1mm} \\
\hspace{-4mm} \begin{sideways} \hspace{-3mm} \tiny{Diff} \end{sideways} & \hspace{-4.0mm} \includegraphics[align=c, width=0.035\linewidth]{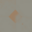} \vspace{1mm} \\
\hspace{-4mm} \begin{sideways} \hspace{-3mm} \tiny{Rough} \end{sideways} & \hspace{-4.0mm} \includegraphics[align=c, width=0.035\linewidth]{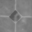} \vspace{1mm} \\
\hspace{-4mm} \begin{sideways} \hspace{-3mm} \tiny{Specu} \end{sideways} & \hspace{-4.0mm} \includegraphics[align=c, width=0.035\linewidth]{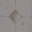} \vspace{1mm} \\
\end{tabular} & \begin{tabular} {cccc}
\hspace{-4mm} \begin{sideways} \hspace{-3mm} \tiny{Normals} \end{sideways} & \hspace{-4.0mm} \includegraphics[align=c, width=0.15\linewidth]{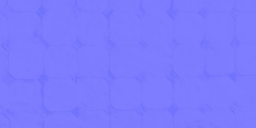} & \hspace{-4.0mm} \includegraphics[align=c, width=0.15\linewidth]{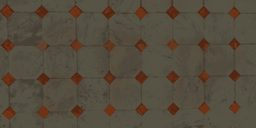} & \hspace{-4mm} \begin{sideways} \hspace{-3mm} \tiny{Diffuse} \end{sideways} \vspace{1mm} \\
\hspace{-4mm} \begin{sideways} \hspace{-3mm} \tiny{Roughness} \end{sideways} & \hspace{-4.0mm} \includegraphics[align=c, width=0.15\linewidth]{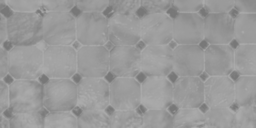} & \hspace{-4.0mm} \includegraphics[align=c, width=0.15\linewidth]{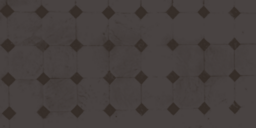} & \hspace{-4mm} \begin{sideways} \hspace{-3mm} \tiny{Specular} \end{sideways} \vspace{1mm} \\
\end{tabular} & \hspace{-4.0mm} \includegraphics[align=c, width=0.16\linewidth]{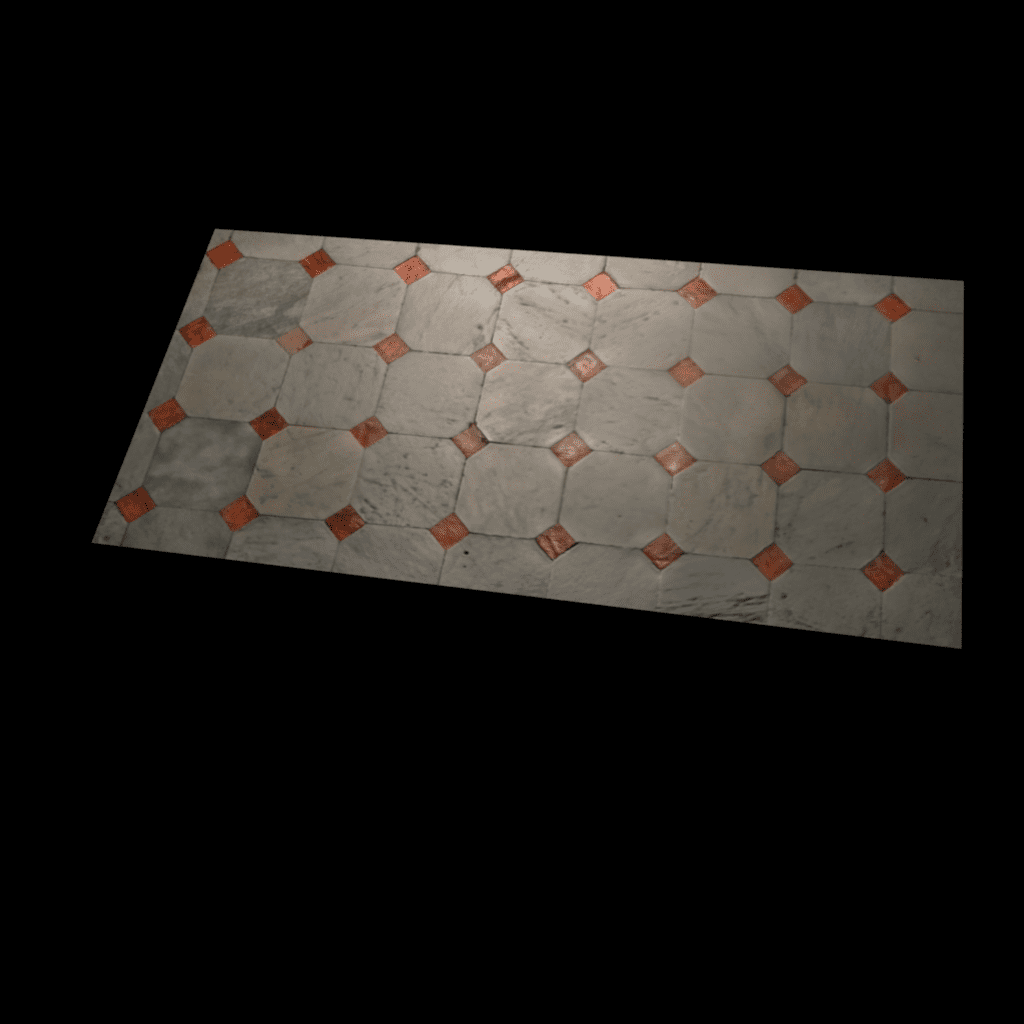} \vspace{1mm} \\
\hspace{-4.0mm} \includegraphics[align=c, width=0.32\linewidth]{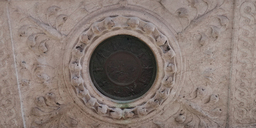} & \begin{tabular} {ccc}
\hspace{-4mm} \begin{sideways} \hspace{-3mm} \tiny{Norm} \end{sideways} & \hspace{-4.0mm} \includegraphics[align=c, width=0.035\linewidth]{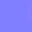} & \hspace{-4.0mm} \includegraphics[align=c, width=0.035\linewidth]{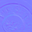} \vspace{1mm} \\
\hspace{-4mm} \begin{sideways} \hspace{-3mm} \tiny{Diff} \end{sideways} & \hspace{-4.0mm} \includegraphics[align=c, width=0.035\linewidth]{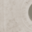} & \hspace{-4.0mm} \includegraphics[align=c, width=0.035\linewidth]{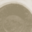} \vspace{1mm} \\
\hspace{-4mm} \begin{sideways} \hspace{-3mm} \tiny{Rough} \end{sideways} & \hspace{-4.0mm} \includegraphics[align=c, width=0.035\linewidth]{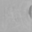} & \hspace{-4.0mm} \includegraphics[align=c, width=0.035\linewidth]{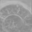} \vspace{1mm} \\
\hspace{-4mm} \begin{sideways} \hspace{-3mm} \tiny{Specu} \end{sideways} & \hspace{-4.0mm} \includegraphics[align=c, width=0.035\linewidth]{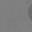} & \hspace{-4.0mm} \includegraphics[align=c, width=0.035\linewidth]{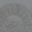} \vspace{1mm} \\
\end{tabular} & \begin{tabular} {cccc}
\hspace{-4mm} \begin{sideways} \hspace{-3mm} \tiny{Normals} \end{sideways} & \hspace{-4.0mm} \includegraphics[align=c, width=0.15\linewidth]{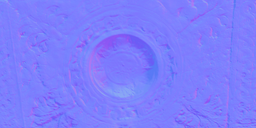} & \hspace{-4.0mm} \includegraphics[align=c, width=0.15\linewidth]{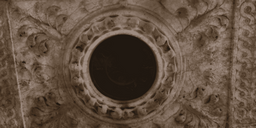} & \hspace{-4mm} \begin{sideways} \hspace{-3mm} \tiny{Diffuse} \end{sideways} \vspace{1mm} \\
\hspace{-4mm} \begin{sideways} \hspace{-3mm} \tiny{Roughness} \end{sideways} & \hspace{-4.0mm} \includegraphics[align=c, width=0.15\linewidth]{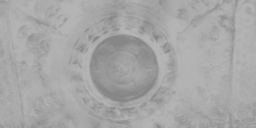} & \hspace{-4.0mm} \includegraphics[align=c, width=0.15\linewidth]{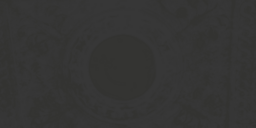} & \hspace{-4mm} \begin{sideways} \hspace{-3mm} \tiny{Specular} \end{sideways} \vspace{1mm} \\
\end{tabular} & \hspace{-4.0mm} \includegraphics[align=c, width=0.16\linewidth]{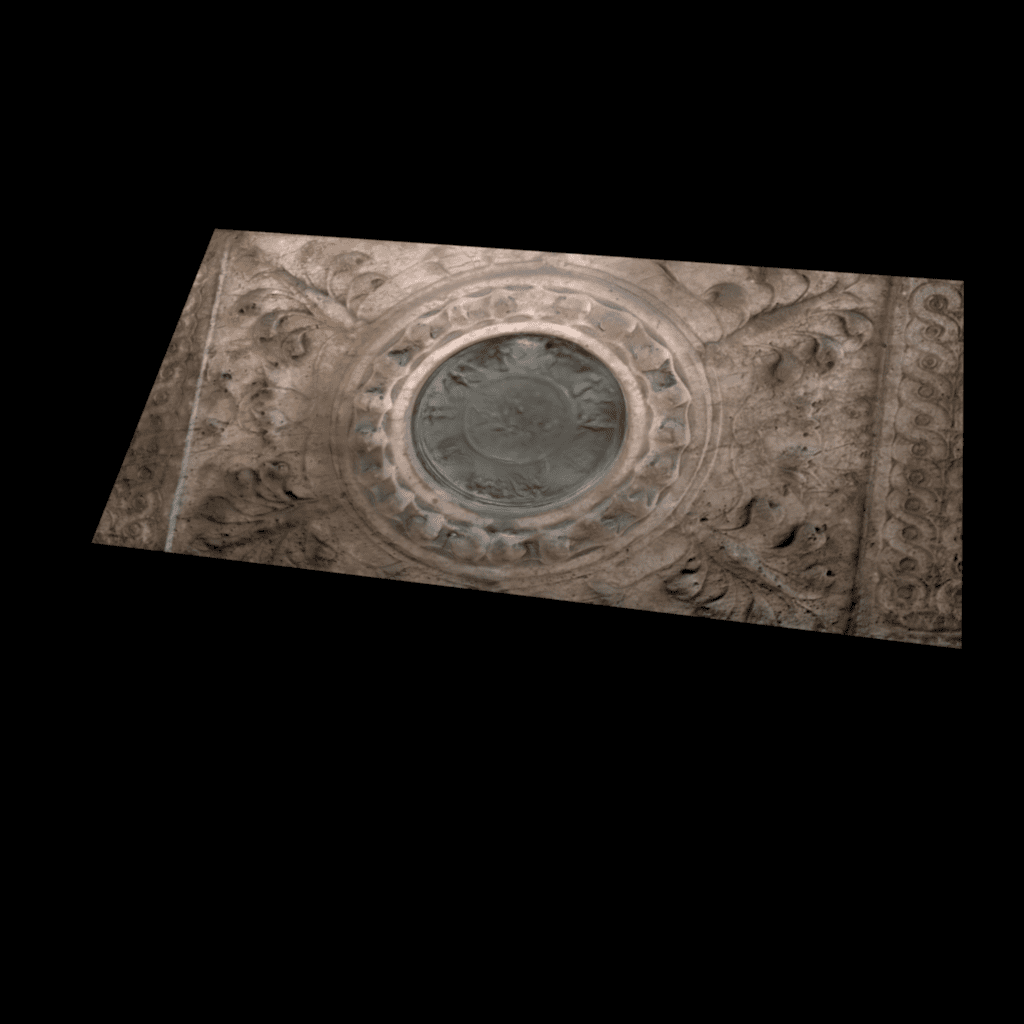} \vspace{1mm} \\
\end{tabular}
\caption{Real-world surface captured on-site with our method. We used a single flash picture to capture the shininess of the tiles, which is propagated to all tiles of the large floor. 
We used two flash pictures for the second example, one for the diffuse stone and the other one for the more shiny metal disk. 
Please zoom on the .pdf to appreciate the high-resolution details of the individual SVBRDF maps. Images of resolution $2048\times1024$.}
\label{fig:realResults}
\end{figure*}

\begin{figure*}
\begin{tabular} {cccc}
HD input picture & Examplars & Results & Rendering \vspace{1mm} \\
\hspace{-4.0mm} \includegraphics[align=c, width=0.3\linewidth]{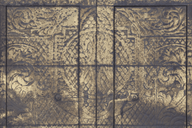} & \begin{tabular} {ccc}
\hspace{-4mm} \begin{sideways} \hspace{-3mm} \tiny{Diff} \end{sideways} & \hspace{-4.0mm} \includegraphics[align=c, width=0.045\linewidth]{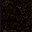} & \hspace{-4.0mm} \includegraphics[align=c, width=0.045\linewidth]{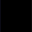} \vspace{1mm} \\
\hspace{-4mm} \begin{sideways} \hspace{-3mm} \tiny{Rough} \end{sideways} & \hspace{-4.0mm} \includegraphics[align=c, width=0.045\linewidth]{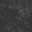} & \hspace{-4.0mm} \includegraphics[align=c, width=0.045\linewidth]{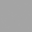} \vspace{1mm} \\
\hspace{-4mm} \begin{sideways} \hspace{-3mm} \tiny{Specu} \end{sideways} & \hspace{-4.0mm} \includegraphics[align=c, width=0.045\linewidth]{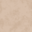} & \hspace{-4.0mm} \includegraphics[align=c, width=0.045\linewidth]{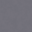} \vspace{1mm} \\
\end{tabular} & \begin{tabular} {cccc}
\hspace{-4mm} \begin{sideways} \hspace{-3mm} \tiny{Normals} \end{sideways} & \hspace{-4.0mm} \includegraphics[align=c, width=0.14\linewidth]{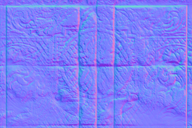} & \hspace{-4.0mm} \includegraphics[align=c, width=0.14\linewidth]{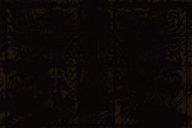} & \hspace{-4mm} \begin{sideways} \hspace{-3mm} \tiny{Diffuse} \end{sideways} \vspace{1mm} \\
\hspace{-4mm} \begin{sideways} \hspace{-3mm} \tiny{Roughness} \end{sideways} & \hspace{-4.0mm} \includegraphics[align=c, width=0.14\linewidth]{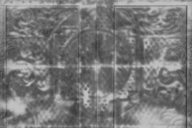} & \hspace{-4.0mm} \includegraphics[align=c, width=0.14\linewidth]{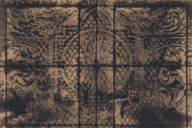} & \hspace{-4mm} \begin{sideways} \hspace{-3mm} \tiny{Specular} \end{sideways} \vspace{1mm} \\
\end{tabular} & \hspace{-4.0mm} \includegraphics[align=c, width=0.2\linewidth]{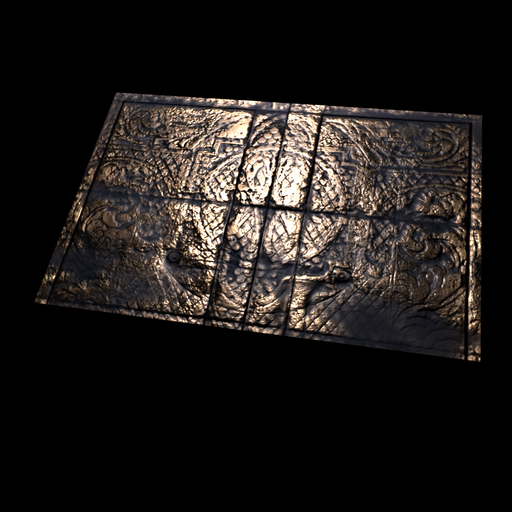} \vspace{1mm} \\
\hspace{-4.0mm} \includegraphics[align=c, width=0.3\linewidth]{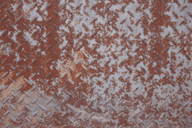} & \begin{tabular} {ccc}
\hspace{-4mm} \begin{sideways} \hspace{-3mm} \tiny{Norm} \end{sideways} & \hspace{-4.0mm} \includegraphics[align=c, width=0.045\linewidth]{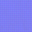} & \hspace{-4.0mm} \includegraphics[align=c, width=0.045\linewidth]{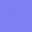} \vspace{1mm} \\
\hspace{-4mm} \begin{sideways} \hspace{-3mm} \tiny{Diff} \end{sideways} & \hspace{-4.0mm} \includegraphics[align=c, width=0.045\linewidth]{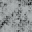} & \hspace{-4.0mm} \includegraphics[align=c, width=0.045\linewidth]{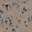} \vspace{1mm} \\
\hspace{-4mm} \begin{sideways} \hspace{-3mm} \tiny{Rough} \end{sideways} & \hspace{-4.0mm} \includegraphics[align=c, width=0.045\linewidth]{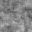} & \hspace{-4.0mm} \includegraphics[align=c, width=0.045\linewidth]{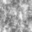} \vspace{1mm} \\
\hspace{-4mm} \begin{sideways} \hspace{-3mm} \tiny{Specu} \end{sideways} & \hspace{-4.0mm} \includegraphics[align=c, width=0.045\linewidth]{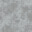} & \hspace{-4.0mm} \includegraphics[align=c, width=0.045\linewidth]{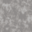} \vspace{1mm} \\
\end{tabular} & \begin{tabular} {cccc}
\hspace{-4mm} \begin{sideways} \hspace{-3mm} \tiny{Normals} \end{sideways} & \hspace{-4.0mm} \includegraphics[align=c, width=0.14\linewidth]{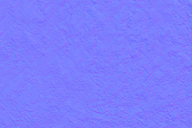} & \hspace{-4.0mm} \includegraphics[align=c, width=0.14\linewidth]{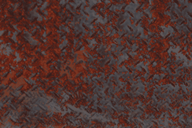} & \hspace{-4mm} \begin{sideways} \hspace{-3mm} \tiny{Diffuse} \end{sideways} \vspace{1mm} \\
\hspace{-4mm} \begin{sideways} \hspace{-3mm} \tiny{Roughness} \end{sideways} & \hspace{-4.0mm} \includegraphics[align=c, width=0.14\linewidth]{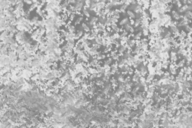} & \hspace{-4.0mm} \includegraphics[align=c, width=0.14\linewidth]{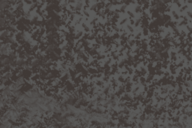} & \hspace{-4mm} \begin{sideways} \hspace{-3mm} \tiny{Specular} \end{sideways} \vspace{1mm} \\
\end{tabular} & \hspace{-4.0mm} \includegraphics[align=c, width=0.2\linewidth]{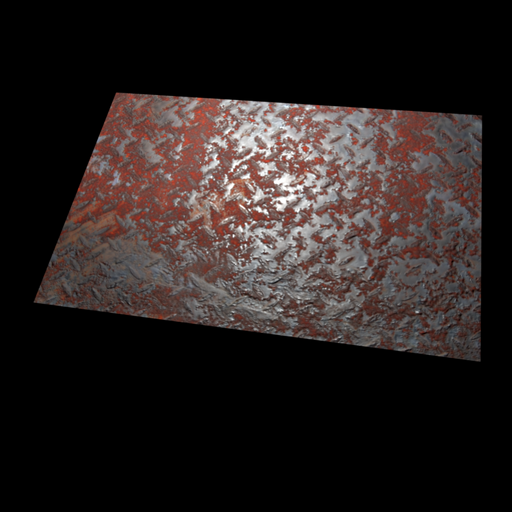} \vspace{1mm} \\
\hspace{-4.0mm} \includegraphics[align=c, width=0.3\linewidth]{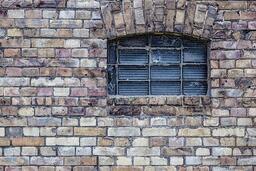} & \begin{tabular} {cc}
\hspace{-4mm} \begin{sideways} \hspace{-3mm} \tiny{Norm} \end{sideways} & \hspace{-4.0mm} \includegraphics[align=c, width=0.045\linewidth]{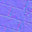} \vspace{1mm} \\
\hspace{-4mm} \begin{sideways} \hspace{-3mm} \tiny{Diff} \end{sideways} & \hspace{-4.0mm} \includegraphics[align=c, width=0.045\linewidth]{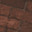} \vspace{1mm} \\
\hspace{-4mm} \begin{sideways} \hspace{-3mm} \tiny{Rough} \end{sideways} & \hspace{-4.0mm} \includegraphics[align=c, width=0.045\linewidth]{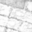} \vspace{1mm} \\
\hspace{-4mm} \begin{sideways} \hspace{-3mm} \tiny{Specu} \end{sideways} & \hspace{-4.0mm} \includegraphics[align=c, width=0.045\linewidth]{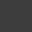} \vspace{1mm} \\
\end{tabular} & \begin{tabular} {cccc}
\hspace{-4mm} \begin{sideways} \hspace{-3mm} \tiny{Normals} \end{sideways} & \hspace{-4.0mm} \includegraphics[align=c, width=0.14\linewidth]{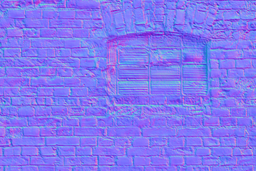} & \hspace{-4.0mm} \includegraphics[align=c, width=0.14\linewidth]{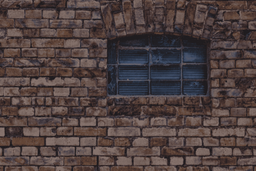} & \hspace{-4mm} \begin{sideways} \hspace{-3mm} \tiny{Diffuse} \end{sideways} \vspace{1mm} \\
\hspace{-4mm} \begin{sideways} \hspace{-3mm} \tiny{Roughness} \end{sideways} & \hspace{-4.0mm} \includegraphics[align=c, width=0.14\linewidth]{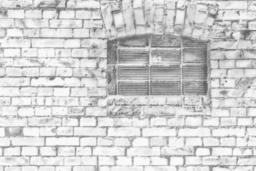} & \hspace{-4.0mm} \includegraphics[align=c, width=0.14\linewidth]{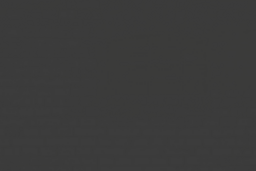} & \hspace{-4mm} \begin{sideways} \hspace{-3mm} \tiny{Specular} \end{sideways} \vspace{1mm} \\
\end{tabular} & \hspace{-4.0mm} \includegraphics[align=c, width=0.2\linewidth]{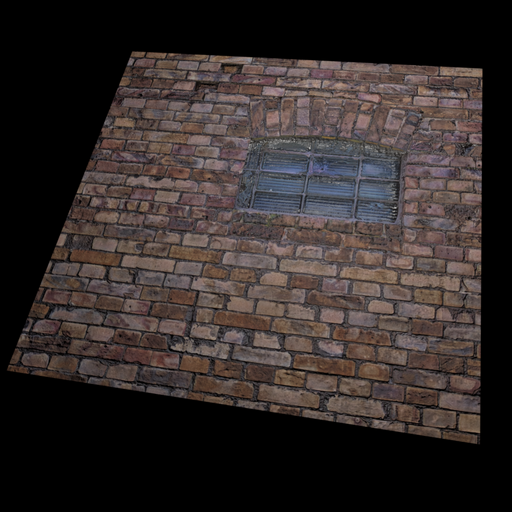} \vspace{1mm} \\

\end{tabular}
\caption{Various SVBRDFs estimated from internet images. We selected artist-designed SVBRDF patches as exemplars for gold, paint, rust and bricks. Note how the shiny gold is well transferred to the yellow parts of the top panel, and how the diffuse rust is transferred to the brown parts of the middle plate. Note also that our method produces a plausible interpretation of the window (third row), even though the provided exemplar only contains bricks. Please zoom on the document to appreciate the high-resolution details of the individual SVBRDF maps. Images of resolution $1536\times1024$.}
\label{fig:onlineResults}
\end{figure*}

\section{Evaluation}
\label{sec:eval}
We first present results obtained by applying our method on our own photographs as well as on internet images. We then evaluate the impact of our fine-tuning and data augmentation strategies. Finally, we compare our method with alternative approaches on synthetic data for which we have ground truth SVBRDF maps. Please see supplemental materials for high-resolution SVBRDF parameter maps and animated renderings of all our results. We will release our code and data upon acceptance to ease reproduction.

\subsection{Results}
Our research was originally motivated by the need to quickly acquire the appearance of large-scale surfaces with minimal hardware. Following this first usage scenario, we used a smartphone to photograph a variety of planar objects. For each object, we first captured a single photograph of the entire surface under ambient lighting. We then captured 1-3 close-up flash photos of parts that exhibit characteristic material features. Finally, we ran the single-image network~\cite{Deschaintre18} to obtain SVBRDF exemplars for each close-up. Fig.~\ref{fig:teaser} and \ref{fig:realResults} show a mosaic, tiled floors, and a sculpted wall captured on-site with this approach. Thanks to the exemplars provided, our method faithfully reproduces the varying shininess of the different tiles, and distinguishes rough stone from metal. 

A second usage scenario of our method is to estimate the SVBRDF maps of existing pictures, using pre-designed SVBRDFs as exemplars of similar materials. Fig.~\ref{fig:onlineResults} shows this on three internet images, processed with exemplars from libraries of artist-created procedural SVBRDFs \cite{Substance,CC0Textures}. Our method transfers diffuse and specular reflectances of the exemplars across the surface while conforming to the input image. 
In this workflow, the user selects exemplars that correspond to the materials they would like to see over the large surface. For instance, by choosing appropriate exemplars, the golden part of the mural is successfully interpreted as having low roughness and yellow specular components, and rust is interpreted as a rough orange material. 
The last row of Fig.~\ref{fig:onlineResults} illustrates the behavior of our method when part of the image is not covered by the provided exemplars. In this result, the exemplar guides the interpretation of the bricks, but not of the window. Nevertheless, our method also benefits from generic priors on material appearance learned during pre-training, here to interpret the dark window as more shiny than the brick.

While pre-designed SVBRDFs provide convincing material parameters, many come with normal maps that are either flat or weakly correlated to the target pictures. When this is the case, we ignore the normal map produced by the fine-tuned network and use the one produced by our pre-trained network instead. All results for which the exemplar normal map is not shown were obtained with this approach. 

\begin{figure*}
\begin{tabular} {ccccccc}
\hspace{-2.0mm}HD input picture & \hspace{-2.0mm}Examplars & \hspace{-2.0mm}Normals & \hspace{-2.0mm}Diffuse & \hspace{-2.0mm}Roughness & \hspace{-2.0mm}Specular & \hspace{-2.0mm}Rendering \vspace{1mm} \\

\hspace{-4.0mm} \includegraphics[align=c, width=0.08\linewidth]{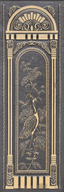} & \begin{tabular} {ccc}
\hspace{-4mm} \begin{sideways} \hspace{-3mm} \tiny{Diffuse} \end{sideways} & \hspace{-4.0mm} \includegraphics[align=c, width=0.05\linewidth]{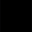} & \hspace{-4.0mm} \includegraphics[align=c, width=0.05\linewidth]{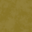} \vspace{1mm} \\
\hspace{-4mm} \begin{sideways} \hspace{-3mm} \tiny{Rough} \end{sideways} & \hspace{-4.0mm} \includegraphics[align=c, width=0.05\linewidth]{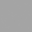} & \hspace{-4.0mm} \includegraphics[align=c, width=0.05\linewidth]{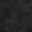} \vspace{1mm} \\
\hspace{-4mm} \begin{sideways} \hspace{-3mm} \tiny{Specular} \end{sideways} & \hspace{-4.0mm} \includegraphics[align=c, width=0.05\linewidth]{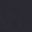} & \hspace{-4.0mm} \includegraphics[align=c, width=0.05\linewidth]{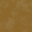} \vspace{1mm} \\
\end{tabular} & \hspace{-4.0mm} \includegraphics[align=c, width=0.08\linewidth]{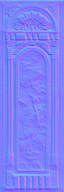} & \hspace{-4.0mm} \includegraphics[align=c, width=0.08\linewidth]{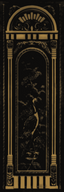} & \hspace{-4.0mm} \includegraphics[align=c, width=0.08\linewidth]{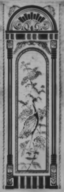} & \hspace{-4.0mm} \includegraphics[align=c, width=0.08\linewidth]{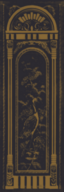} & \hspace{-4.0mm} \includegraphics[align=c, width=0.24\linewidth]{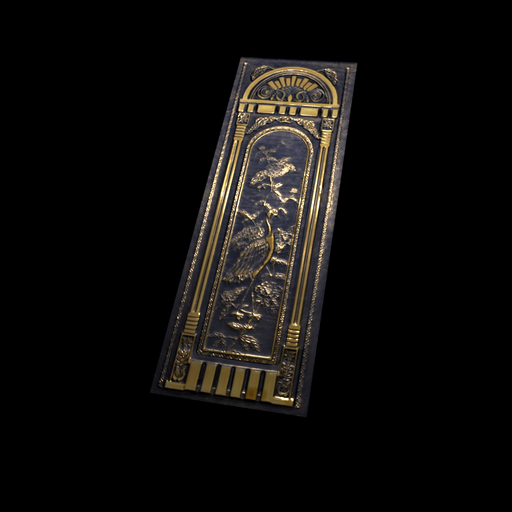} \vspace{1mm} \\

\hspace{-4.0mm} \includegraphics[align=c, width=0.08\linewidth]{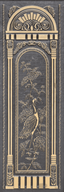} & \begin{tabular} {ccc}
\hspace{-4mm} \begin{sideways} \hspace{-3mm} \tiny{Diffuse} \end{sideways} & \hspace{-4.0mm} \includegraphics[align=c, width=0.05\linewidth]{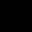} & \hspace{-4.0mm} \includegraphics[align=c, width=0.05\linewidth]{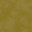} \vspace{1mm} \\
\hspace{-4mm} \begin{sideways} \hspace{-3mm} \tiny{Rough} \end{sideways} & \hspace{-4.0mm} \includegraphics[align=c, width=0.05\linewidth]{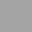} & \hspace{-4.0mm} \includegraphics[align=c, width=0.05\linewidth]{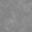} \vspace{1mm} \\
\hspace{-4mm} \begin{sideways} \hspace{-3mm} \tiny{Specular} \end{sideways} & \hspace{-4.0mm} \includegraphics[align=c, width=0.05\linewidth]{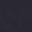} & \hspace{-4.0mm} \includegraphics[align=c, width=0.05\linewidth]{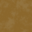} \vspace{1mm} \\
\end{tabular} & \hspace{-4.0mm} \includegraphics[align=c, width=0.08\linewidth]{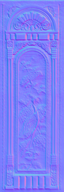} & \hspace{-4.0mm} \includegraphics[align=c, width=0.08\linewidth]{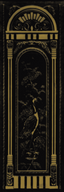} & \hspace{-4.0mm} \includegraphics[align=c, width=0.08\linewidth]{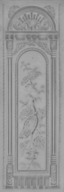} & \hspace{-4.0mm} \includegraphics[align=c, width=0.08\linewidth]{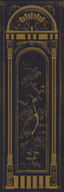} & \hspace{-4.0mm} \includegraphics[align=c, width=0.24\linewidth]{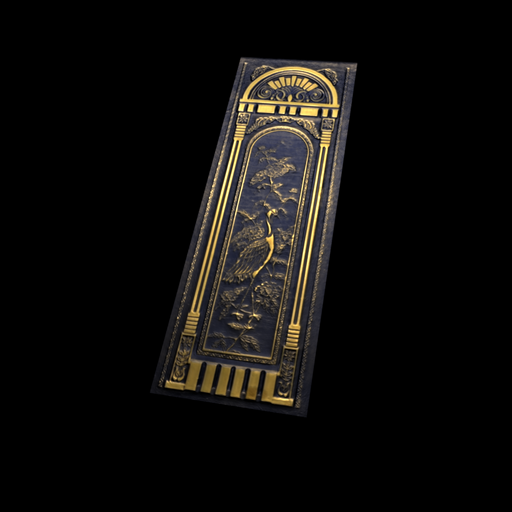} \vspace{1mm} \\

\hspace{-4.0mm} \includegraphics[align=c, width=0.08\linewidth]{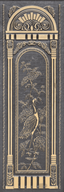} & \begin{tabular} {ccc}
\hspace{-4mm} \begin{sideways} \hspace{-3mm} \tiny{Diffuse} \end{sideways} & \hspace{-4.0mm} \includegraphics[align=c, width=0.05\linewidth]{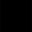} & \hspace{-4.0mm} \includegraphics[align=c, width=0.05\linewidth]{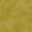} \vspace{1mm} \\
\hspace{-4mm} \begin{sideways} \hspace{-3mm} \tiny{Rough} \end{sideways} & \hspace{-4.0mm} \includegraphics[align=c, width=0.05\linewidth]{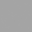} & \hspace{-4.0mm} \includegraphics[align=c, width=0.05\linewidth]{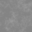} \vspace{1mm} \\
\hspace{-4mm} \begin{sideways} \hspace{-3mm} \tiny{Specular} \end{sideways} & \hspace{-4.0mm} \includegraphics[align=c, width=0.05\linewidth]{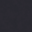} & \hspace{-4.0mm} \includegraphics[align=c, width=0.05\linewidth]{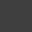} \vspace{1mm} \\
\end{tabular} & \hspace{-4.0mm} \includegraphics[align=c, width=0.08\linewidth]{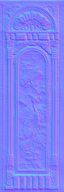} & \hspace{-4.0mm} \includegraphics[align=c, width=0.08\linewidth]{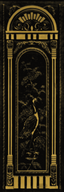} & \hspace{-4.0mm} \includegraphics[align=c, width=0.08\linewidth]{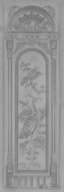} & \hspace{-4.0mm} \includegraphics[align=c, width=0.08\linewidth]{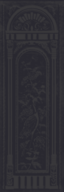} & \hspace{-4.0mm} \includegraphics[align=c, width=0.24\linewidth]{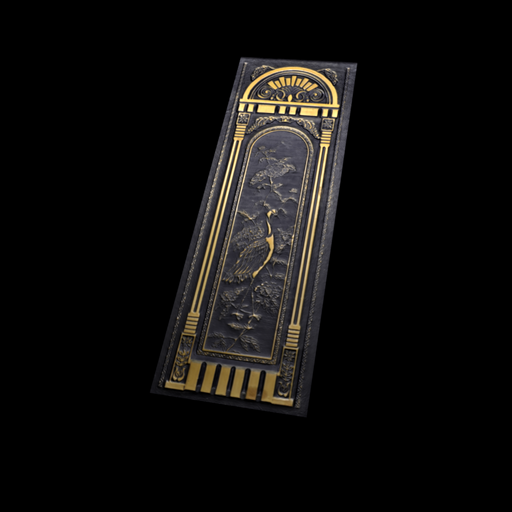} \vspace{1mm} \\

\hspace{-4.0mm} \includegraphics[align=c, width=0.08\linewidth]{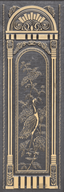} & \begin{tabular} {cccc}
\hspace{-4mm} \begin{sideways} \hspace{-3mm} \tiny{Diffuse} \end{sideways} & \hspace{-4.0mm} \includegraphics[align=c, width=0.05\linewidth]{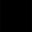} & \hspace{-4.0mm} \includegraphics[align=c, width=0.05\linewidth]{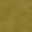} & \hspace{-4.0mm} \includegraphics[align=c, width=0.05\linewidth]{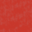} \vspace{1mm} \\
\hspace{-4mm} \begin{sideways} \hspace{-3mm} \tiny{Rough} \end{sideways} & \hspace{-4.0mm} \includegraphics[align=c, width=0.05\linewidth]{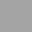} & \hspace{-4.0mm} \includegraphics[align=c, width=0.05\linewidth]{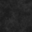} & \hspace{-4.0mm} \includegraphics[align=c, width=0.05\linewidth]{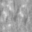} \vspace{1mm} \\
\hspace{-4mm} \begin{sideways} \hspace{-3mm} \tiny{Specular} \end{sideways} & \hspace{-4.0mm} \includegraphics[align=c, width=0.05\linewidth]{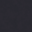} & \hspace{-4.0mm} \includegraphics[align=c, width=0.05\linewidth]{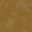} & \hspace{-4.0mm} \includegraphics[align=c, width=0.05\linewidth]{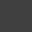} \vspace{1mm} \\
\end{tabular} & \hspace{-4.0mm} \includegraphics[align=c, width=0.08\linewidth]{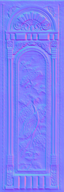} & \hspace{-4.0mm} \includegraphics[align=c, width=0.08\linewidth]{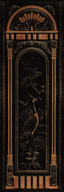} & \hspace{-4.0mm} \includegraphics[align=c, width=0.08\linewidth]{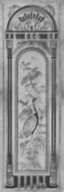} & \hspace{-4.0mm} \includegraphics[align=c, width=0.08\linewidth]{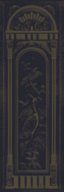} & \hspace{-4.0mm} \includegraphics[align=c, width=0.24\linewidth]{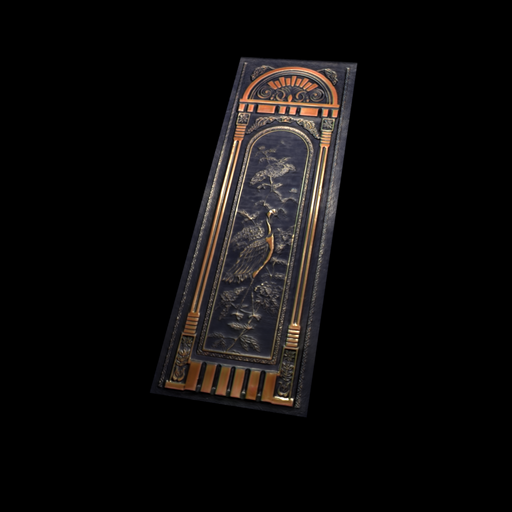} \vspace{1mm} \\

\end{tabular}
\caption{Given the same input picture, we achieve different outcomes by changing the exemplars.
In the first row, we provide an exemplar of a black diffuse material and an exemplar of a shiny yellow metal, which are successfully transferred to the dark and golden parts of the input picture respectively. In the second row, we increased the roughness of the yellow metallic exemplar, which is again successfully propagated to the golden parts of the input. In the third row, we replaced the metallic exemplar by a yellow diffuse material, which results in a SVBRDF where only the diffuse map contains yellow information. Finally, in the fourth row, we included an outlier red diffuse exemplar, which our method tends to mix with the yellow metal to produce a slightly orange diffuse map and a weaker specular map. Images of resolution $512\times1536$.}
\label{fig:exampleInfluence}

\end{figure*}

Fig.~\ref{fig:exampleInfluence} further demonstrates the control that the input exemplars provide on the output SVBRDF. The input picture contains dark and yellow pixels with little in terms of visual cues of their respective shininess. We first selected a dark diffuse and a yellow metallic exemplar to achieve a golden appearance. We next show how changing the exemplar allows us to increase the roughness of the gold, or even to interpret the yellow pixels as diffuse paint. Finally, we also show how our method behaves in the presence of an outlier exemplar, which in this case gives a slight orange tint to the yellow pixels.

\begin{figure*}
\begin{overpic}[scale=1.]{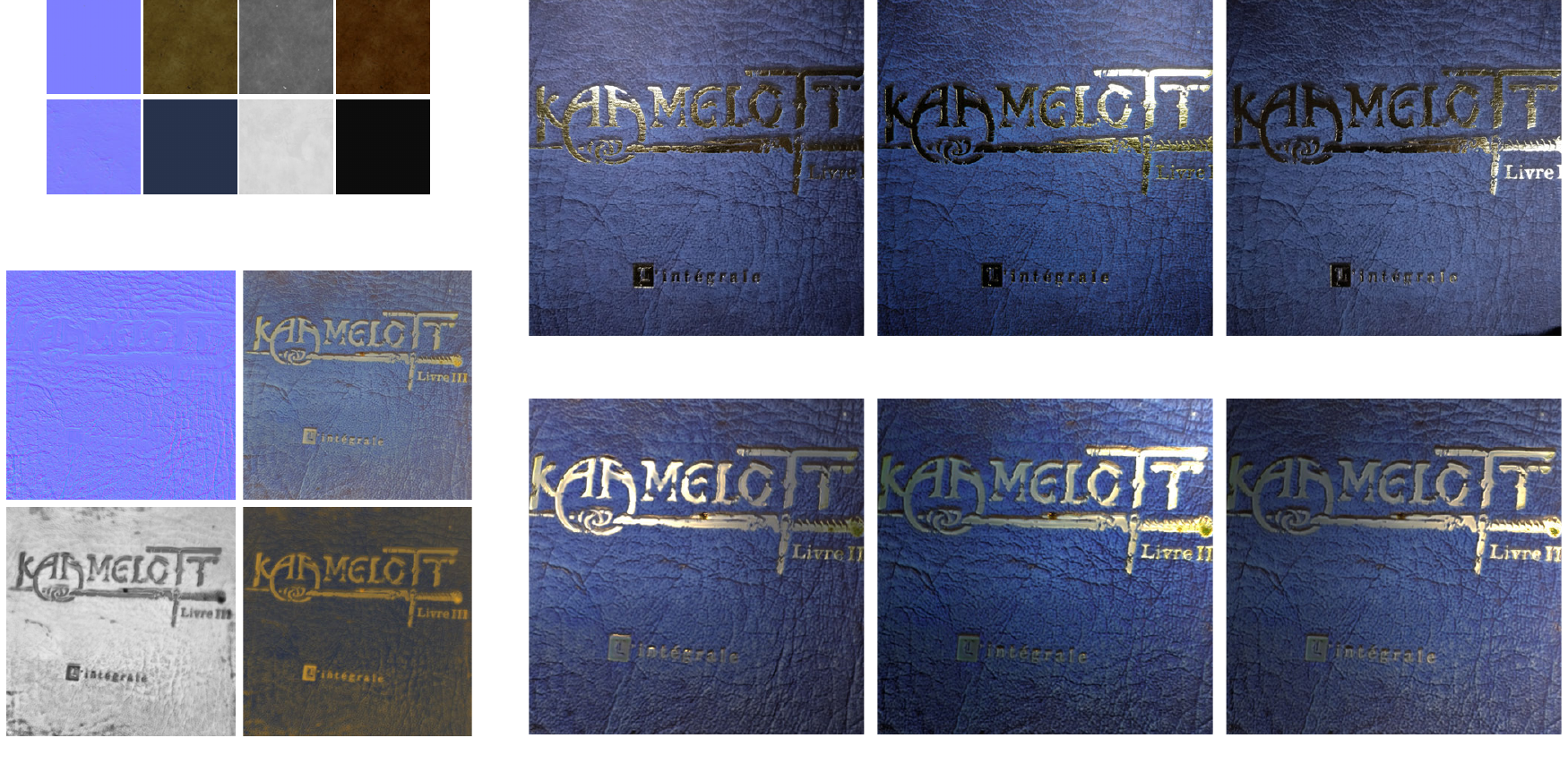}
\put(9,35.5){\small{SVBRDF Exemplars}}
\put(9.5,1){\small{Predicted SVBRDF}}
\put(63.5,26.5){\small{Photographs}}
\put(63.5,1){\small{Renderings}}
\end{overpic}
\vspace{-2mm}
	\caption{Comparison to real-world photographs. We reproduced the appearance of a book cover using a single picture captured under environment lighting, and two exemplars of blue leather and golden material. The top row shows real-world pictures of the book under varying lighting, and the bottom row shows our renderings under similar lighting. A comparison with exemplars obtained with \cite{Deschaintre18} and no exemplars is provided in supplemental material.}
\label{fig:reproductionRealWorld}
\end{figure*}

We show in Fig.~\ref{fig:reproductionRealWorld} a visual comparison between real photographs of a surface and renderings of the SVBRDF created with our method. We used artist-designed SVBRDFs as exemplars for this comparison because the single-image method of Deschaintre et al.~\shortcite{Deschaintre18} fails to recover convincing maps from flash pictures of this complex surface (see supplemental materials for their result). This experiment shows that users can reproduce the desired overall appearance by guiding our method with adequate exemplars.

\begin{figure*}
\begin{tabular} {ccc}
\hspace{-4.0mm} \includegraphics[align=c, width=0.33\linewidth]{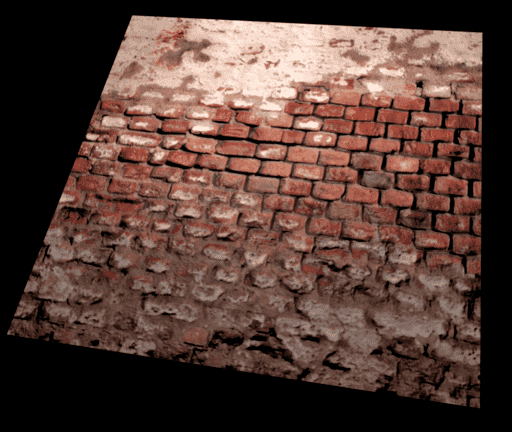} & \hspace{-4.0mm} \includegraphics[align=c, width=0.33\linewidth]{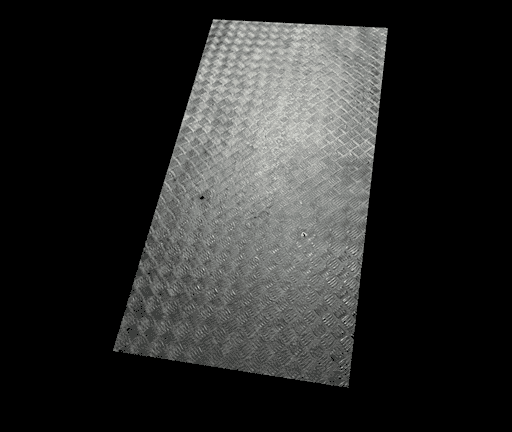} & 
\hspace{-4.0mm} \includegraphics[align=c, width=0.33\linewidth]{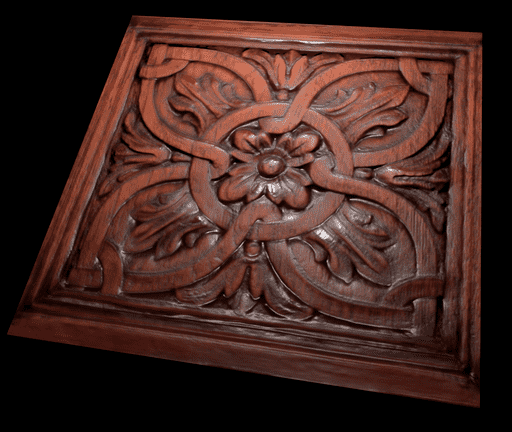} \vspace{1mm} \\
\hspace{-4.0mm} \includegraphics[align=c, width=0.33\linewidth]{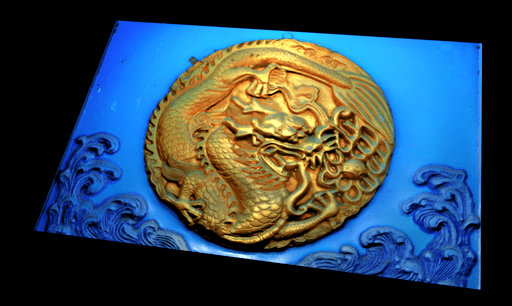} & \hspace{-4.0mm} \includegraphics[align=c, width=0.33\linewidth]{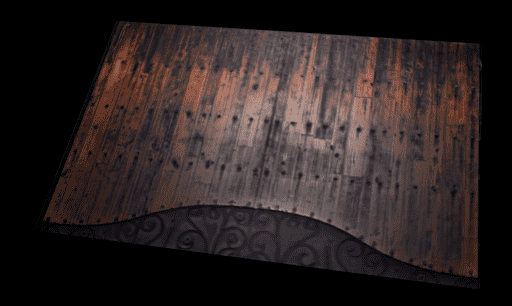} & \hspace{-4.0mm} \includegraphics[align=c, width=0.33\linewidth]{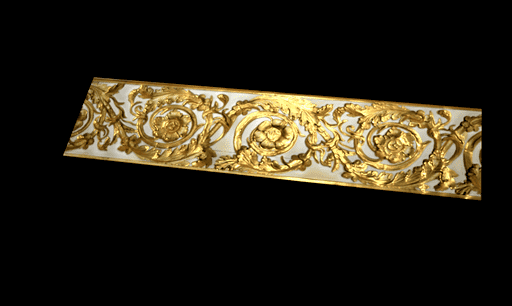} \vspace{1mm} \\
\end{tabular}
\vspace{-2mm}
\caption{A variety of surfaces captured or designed with our method. See supplemental materials for animated renderings.}
\label{fig:moreResults}
\end{figure*}

Finally, Fig.~\ref{fig:moreResults} showcases a variety of SVBRDFs created with our method, either via on-site acquisition or from stock photographs. Note that most of these results represent large, non-square surfaces encoded as high-resolution parameter maps, which contrasts with the small material samples often shown in related work.

\subsection{Ablation study}
\begin{figure*}
\begin{tabular} {ccccc}
Method & HD input picture & Examplars & Results & Rendering \vspace{1mm} \\

\begin{sideways}\hspace{-6mm}No fine-tuning\end{sideways} & \hspace{-4.0mm} \includegraphics[align=c, width=0.2\linewidth]{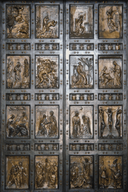} &  & \begin{tabular} {cccc}
\hspace{-4mm} \begin{sideways} \hspace{-4mm} \tiny{Normals} \end{sideways} & \hspace{-4.0mm} \includegraphics[align=c, width=0.095\linewidth]{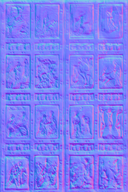} & \hspace{-4.0mm} \includegraphics[align=c, width=0.095\linewidth]{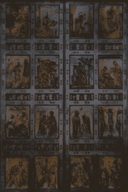} & \hspace{-4mm} \begin{sideways} \hspace{-4mm} \tiny{Diffuse} \end{sideways} \vspace{1mm} \\
\hspace{-4mm} \begin{sideways} \hspace{-4mm} \tiny{Roughness} \end{sideways} & \hspace{-4.0mm} \includegraphics[align=c, width=0.095\linewidth]{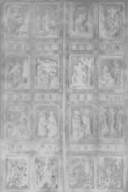} & \hspace{-4.0mm} \includegraphics[align=c, width=0.095\linewidth]{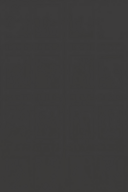} & \hspace{-4mm} \begin{sideways} \hspace{-4mm} \tiny{Specular} \end{sideways} \vspace{1mm} \\
\end{tabular} & \hspace{-4.0mm} \includegraphics[align=c, width=0.29\linewidth]{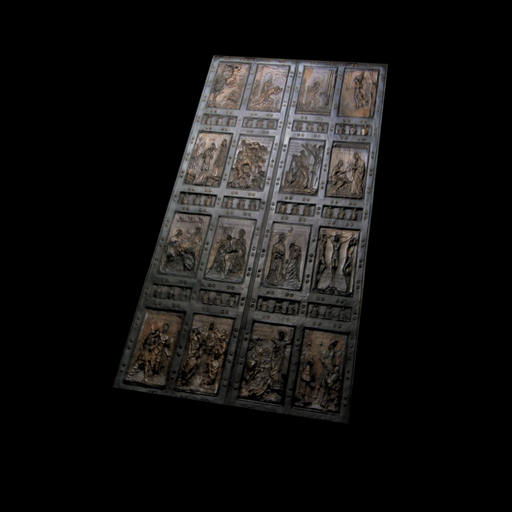} \vspace{1mm} \\

\begin{sideways}\hspace{-10mm}No data augmentation\end{sideways} & \hspace{-4.0mm} \includegraphics[align=c, width=0.2\linewidth]{Figures/ComparisonNoPosttrainingNoPerlin/Inputs/decoratedDoor.png} & \begin{tabular} {ccc}
\hspace{-4mm} \begin{sideways} \hspace{-4mm} \tiny{Diffuse} \end{sideways} & \hspace{-4.0mm} \includegraphics[align=c, width=0.07\linewidth]{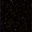} & \hspace{-4.0mm} \includegraphics[align=c, width=0.07\linewidth]{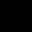} \vspace{1mm} \\
\hspace{-4mm} \begin{sideways} \hspace{-4mm} \tiny{Roughness} \end{sideways} & \hspace{-4.0mm} \includegraphics[align=c, width=0.07\linewidth]{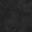} & \hspace{-4.0mm} \includegraphics[align=c, width=0.07\linewidth]{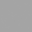} \vspace{1mm} \\
\hspace{-4mm} \begin{sideways} \hspace{-4mm} \tiny{Specular} \end{sideways} & \hspace{-4.0mm} \includegraphics[align=c, width=0.07\linewidth]{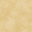} & \hspace{-4.0mm} \includegraphics[align=c, width=0.07\linewidth]{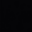} \vspace{1mm} \\
\end{tabular} & \begin{tabular} {cccc}
\hspace{-4mm} \begin{sideways} \hspace{-4mm} \tiny{Normals} \end{sideways} & \hspace{-4.0mm} \includegraphics[align=c, width=0.095\linewidth]{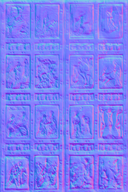} & \hspace{-4.0mm} \includegraphics[align=c, width=0.095\linewidth]{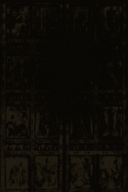} & \hspace{-4mm} \begin{sideways} \hspace{-4mm} \tiny{Diffuse} \end{sideways} \vspace{1mm} \\
\hspace{-4mm} \begin{sideways} \hspace{-4mm} \tiny{Roughness} \end{sideways} & \hspace{-4.0mm} \includegraphics[align=c, width=0.095\linewidth]{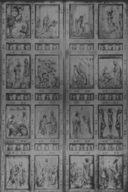} & \hspace{-4.0mm} \includegraphics[align=c, width=0.095\linewidth]{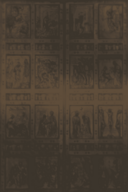} & \hspace{-4mm} \begin{sideways} \hspace{-4mm} \tiny{Specular} \end{sideways} \vspace{1mm} \\
\end{tabular} & \hspace{-4.0mm} \includegraphics[align=c, width=0.29\linewidth]{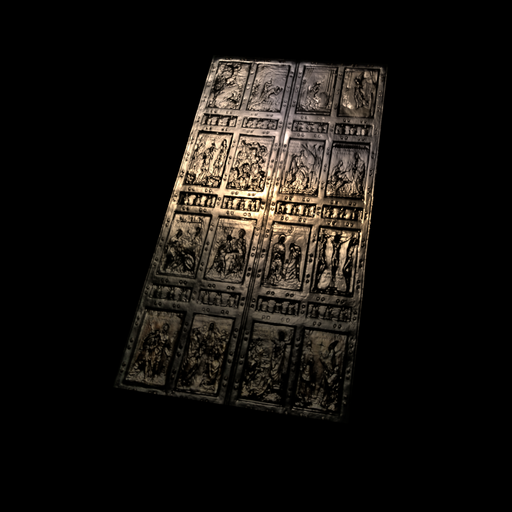} \vspace{1mm} \\

\begin{sideways}\hspace{-2mm}Ours\end{sideways} & \hspace{-4.0mm} \includegraphics[align=c, width=0.2\linewidth]{Figures/ComparisonNoPosttrainingNoPerlin/Inputs/decoratedDoor.png} & \begin{tabular} {ccc}
\hspace{-4mm} \begin{sideways} \hspace{-4mm} \tiny{Diffuse} \end{sideways} & \hspace{-4.0mm} \includegraphics[align=c, width=0.07\linewidth]{Figures/ComparisonNoPosttrainingNoPerlin/Inputs/decoratedDoor_example0_1.png} & \hspace{-4.0mm} \includegraphics[align=c, width=0.07\linewidth]{Figures/ComparisonNoPosttrainingNoPerlin/Inputs/decoratedDoor_example1_1.png} \vspace{1mm} \\
\hspace{-4mm} \begin{sideways} \hspace{-4mm} \tiny{Roughness} \end{sideways} & \hspace{-4.0mm} \includegraphics[align=c, width=0.07\linewidth]{Figures/ComparisonNoPosttrainingNoPerlin/Inputs/decoratedDoor_example0_2.png} & \hspace{-4.0mm} \includegraphics[align=c, width=0.07\linewidth]{Figures/ComparisonNoPosttrainingNoPerlin/Inputs/decoratedDoor_example1_2.png} \vspace{1mm} \\
\hspace{-4mm} \begin{sideways} \hspace{-4mm} \tiny{Specular} \end{sideways} & \hspace{-4.0mm} \includegraphics[align=c, width=0.07\linewidth]{Figures/ComparisonNoPosttrainingNoPerlin/Inputs/decoratedDoor_example0_3.png} & \hspace{-4.0mm} \includegraphics[align=c, width=0.07\linewidth]{Figures/ComparisonNoPosttrainingNoPerlin/Inputs/decoratedDoor_example1_3.png} \vspace{1mm} \\
\end{tabular} & \begin{tabular} {cccc}
\hspace{-4mm} \begin{sideways} \hspace{-4mm} \tiny{Normals} \end{sideways} & \hspace{-4.0mm} \includegraphics[align=c, width=0.095\linewidth]{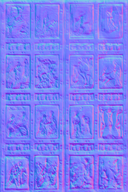} & \hspace{-4.0mm} \includegraphics[align=c, width=0.095\linewidth]{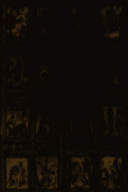} & \hspace{-4mm} \begin{sideways} \hspace{-4mm} \tiny{Diffuse} \end{sideways} \vspace{1mm} \\
\hspace{-4mm} \begin{sideways} \hspace{-4mm} \tiny{Roughness} \end{sideways} & \hspace{-4.0mm} \includegraphics[align=c, width=0.095\linewidth]{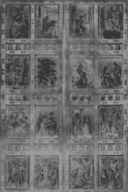} & \hspace{-4.0mm} \includegraphics[align=c, width=0.095\linewidth]{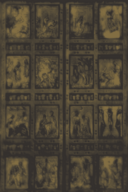} & \hspace{-4mm} \begin{sideways} \hspace{-4mm} \tiny{Specular} \end{sideways} \vspace{1mm} \\
\end{tabular} & \hspace{-4.0mm} \includegraphics[align=c, width=0.29\linewidth]{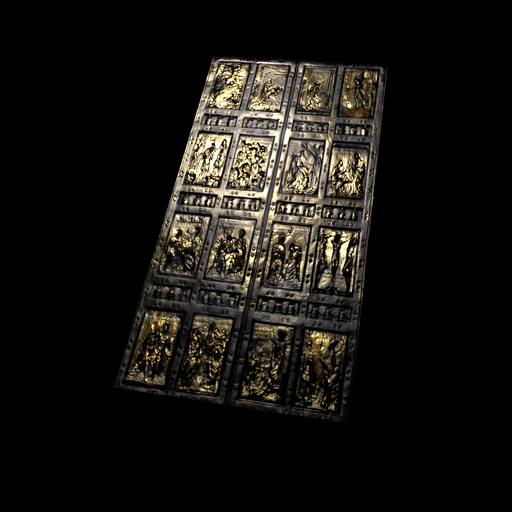} \vspace{1mm} \\
\end{tabular}
\caption{Ablation study. The baseline single-image network of Deschaintre et al. \shortcite{Deschaintre18} interprets this weathered golden door as made of rough plastic (first row). Fine-tuning this network on two exemplars without data augmentation yields a uniform golden appearance (second row). Thanks to data augmentation, our method successfully distinguishes the shiny golden parts from the more diffuse dark parts (third row). See supplemental materials for additional ablation results. Image of resolution $1024\times1536$.}
\label{fig:ComparisonNoPerlinNoPostTraining}
\end{figure*}

We use the single-image network of Deschaintre et al.~\shortcite{Deschaintre18} as a backbone for SVBRDF prediction. Fig.~\ref{fig:ComparisonNoPerlinNoPostTraining} (first row) shows results of their method trained on our dataset of images rendered under random directional lighting. Without additional guidance, this method interprets the weathered golden door as made of rough plastic. Fig.~\ref{fig:ComparisonNoPerlinNoPostTraining} (second row) shows how fine-tuning this single-image network on two exemplars without data augmentation brings a golden appearance but distributes it uniformly over the surface. In our experiments, this tendency to produce uniform maps happens especially when the input exemplars are themselves uniform. By combining the exemplars to form random patterns, our data augmentation helps the transfer of the golden appearance to the least weathered parts of the door (Fig.~\ref{fig:ComparisonNoPerlinNoPostTraining}, third row).


\begin{table*}
\centering
\begin{tabular}{|c|c|c|c|c|c|c|c|}
\hline 
	& \cite{Deschaintre18}&\cite{Li17}&Few shot &Ours& Ours\\ 
	&No Flash& & style transfer &  \cite{Deschaintre18} exemplar & GT exemplar \\
\hline 
Normals & 0.045 &  & 0.043 & 0.04  & \textbf{0.039} \\ 
\hline 
Diffuse & 0.092 &  & 0.095 & 0.059 & \textbf{0.028} \\ 
\hline 
Roughness & 0.215 &  & 0.195 & 0.142 & \textbf{0.056}
\\ 
\hline 
Specular & 0.016 &  & 0.015 & 0.021 & \textbf{0.005}
\\ 
\hline 
Renderings & 0.122 & 0.256 & 0.124 & 0.086 & \textbf{0.071}
\\ 
\hline 
\end{tabular}
	\caption{Numerical comparison to alternative methods using the RMSE metric (smaller is better), performed on synthetic SVBRDFs. Our method outperforms existing single-image algorithms thanks to the guidance of the exemplar (only one exemplar used). We only report the rendering error for \cite{Li17} because this method outputs a different BRDF model than ours.}
\label{tab:rmse}

\end{table*}

\subsection{Comparisons}
To our knowledge, our method is the first to offer by-example guidance for deep SVBRDF inference. We compare to related work on style transfer, as well as to single-image alternatives. We use synthetic SVBRDFs for these comparisons, which allows visual comparison to the ground truth maps, as well as numerical evaluation.

\begin{figure*}
\begin{tabular} {ccccc}
Method & HD input picture & Examplar & Results & Rendering \vspace{1mm} \\
\begin{sideways}\hspace{-2.0mm}AdaIN\end{sideways} & \hspace{-4.0mm} \includegraphics[align=c, width=0.21\linewidth]{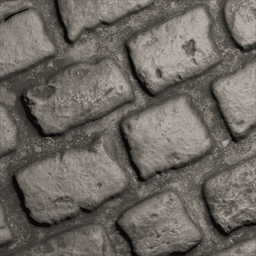} & \begin{tabular} {cc}
\hspace{-4mm} \begin{sideways} \hspace{-3mm} \tiny{Normals} \end{sideways} & \hspace{-4.0mm} \includegraphics[align=c, width=0.045\linewidth]{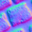} \vspace{1mm} \\
\hspace{-4mm} \begin{sideways} \hspace{-3mm} \tiny{Diffuse} \end{sideways} & \hspace{-4.0mm} \includegraphics[align=c, width=0.045\linewidth]{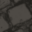} \vspace{1mm} \\
\hspace{-4mm} \begin{sideways} \hspace{-3mm} \tiny{Rough} \end{sideways} & \hspace{-4.0mm} \includegraphics[align=c, width=0.045\linewidth]{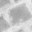} \vspace{1mm} \\
\hspace{-4mm} \begin{sideways} \hspace{-3mm} \tiny{Specu} \end{sideways} & \hspace{-4.0mm} \includegraphics[align=c, width=0.045\linewidth]{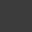} \vspace{1mm} \\
\end{tabular} & \begin{tabular} {cccc}
\hspace{-4mm} \begin{sideways} \hspace{-3mm} \tiny{Normals} \end{sideways} & \hspace{-4.0mm} \includegraphics[align=c, width=0.1\linewidth]{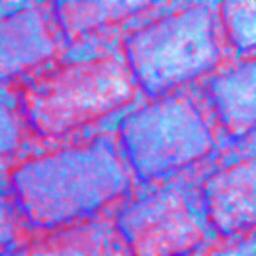} & \hspace{-4.0mm} \includegraphics[align=c, width=0.1\linewidth]{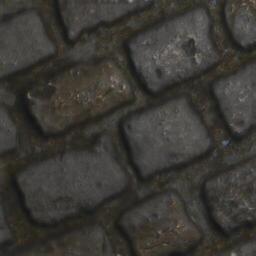} & \hspace{-4mm} \begin{sideways} \hspace{-3mm} \tiny{Diffuse} \end{sideways} \vspace{1mm} \\
\hspace{-4mm} \begin{sideways} \hspace{-3mm} \tiny{Roughness} \end{sideways} & \hspace{-4.0mm} \includegraphics[align=c, width=0.1\linewidth]{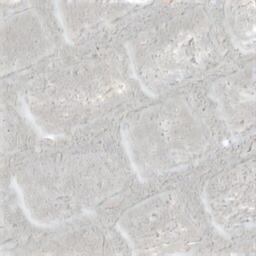} & \hspace{-4.0mm} \includegraphics[align=c, width=0.1\linewidth]{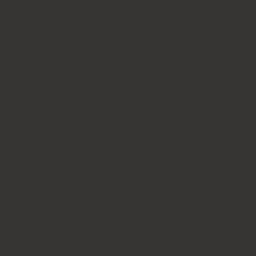} & \hspace{-4mm} \begin{sideways} \hspace{-3mm} \tiny{Specular} \end{sideways} \vspace{1mm} \\
\end{tabular} & \hspace{-4.0mm} \includegraphics[align=c, width=0.21\linewidth]{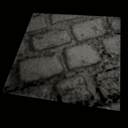} \vspace{1mm} \\
	\begin{sideways}\hspace{-10.0mm}Texture synthesis\end{sideways} & \hspace{-4.0mm} \includegraphics[align=c, width=0.21\linewidth]{Figures/ComparisonAdaINPatchMatch/Inputs/558_brick_uneven_stones.png} & \begin{tabular} {cc}
\hspace{-4mm} \begin{sideways} \hspace{-3mm} \tiny{Normals} \end{sideways} & \hspace{-4.0mm} \includegraphics[align=c, width=0.045\linewidth]{Figures/ComparisonAdaINPatchMatch/Inputs/558_brick_uneven_stones_example_0.png} \vspace{1mm} \\
\hspace{-4mm} \begin{sideways} \hspace{-3mm} \tiny{Diffuse} \end{sideways} & \hspace{-4.0mm} \includegraphics[align=c, width=0.045\linewidth]{Figures/ComparisonAdaINPatchMatch/Inputs/558_brick_uneven_stones_example_1.png} \vspace{1mm} \\
\hspace{-4mm} \begin{sideways} \hspace{-3mm} \tiny{Rough} \end{sideways} & \hspace{-4.0mm} \includegraphics[align=c, width=0.045\linewidth]{Figures/ComparisonAdaINPatchMatch/Inputs/558_brick_uneven_stones_example_2.png} \vspace{1mm} \\
\hspace{-4mm} \begin{sideways} \hspace{-3mm} \tiny{Specu} \end{sideways} & \hspace{-4.0mm} \includegraphics[align=c, width=0.045\linewidth]{Figures/ComparisonAdaINPatchMatch/Inputs/558_brick_uneven_stones_example_3.png} \vspace{1mm} \\
\end{tabular} & \begin{tabular} {cccc}
\hspace{-4mm} \begin{sideways} \hspace{-3mm} \tiny{Normals} \end{sideways} & \hspace{-4.0mm} \includegraphics[align=c, width=0.1\linewidth]{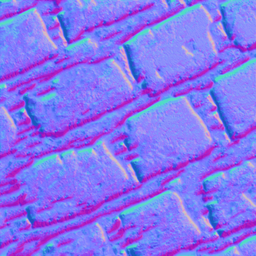} & \hspace{-4.0mm} \includegraphics[align=c, width=0.1\linewidth]{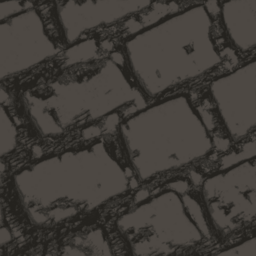} & \hspace{-4mm} \begin{sideways} \hspace{-3mm} \tiny{Diffuse} \end{sideways} \vspace{1mm} \\
\hspace{-4mm} \begin{sideways} \hspace{-3mm} \tiny{Roughness} \end{sideways} & \hspace{-4.0mm} \includegraphics[align=c, width=0.1\linewidth]{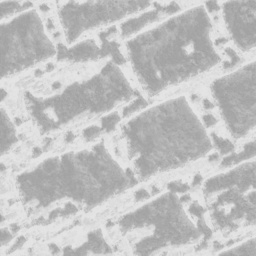} & \hspace{-4.0mm} \includegraphics[align=c, width=0.1\linewidth]{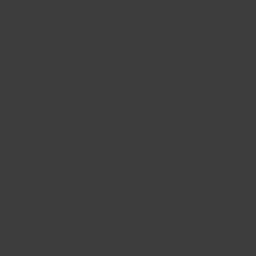} & \hspace{-4mm} \begin{sideways} \hspace{-3mm} \tiny{Specular} \end{sideways} \vspace{1mm} \\
\end{tabular} & \hspace{-4.0mm} \includegraphics[align=c, width=0.21\linewidth]{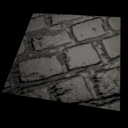} \vspace{1mm} \\
\begin{sideways}GT\end{sideways} & \hspace{-4.0mm} \includegraphics[align=c, width=0.21\linewidth]{Figures/ComparisonAdaINPatchMatch/Inputs/558_brick_uneven_stones.png} &  & \begin{tabular} {cccc}
\hspace{-4mm} \begin{sideways} \hspace{-3mm} \tiny{Normals} \end{sideways} & \hspace{-4.0mm} \includegraphics[align=c, width=0.1\linewidth]{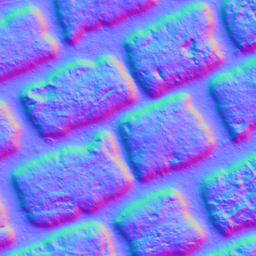} & \hspace{-4.0mm} \includegraphics[align=c, width=0.1\linewidth]{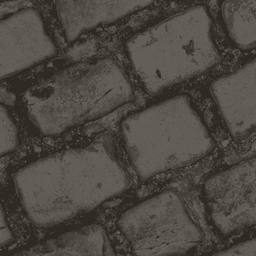} & \hspace{-4mm} \begin{sideways} \hspace{-3mm} \tiny{Diffuse} \end{sideways} \vspace{1mm} \\
\hspace{-4mm} \begin{sideways} \hspace{-3mm} \tiny{Roughness} \end{sideways} & \hspace{-4.0mm} \includegraphics[align=c, width=0.1\linewidth]{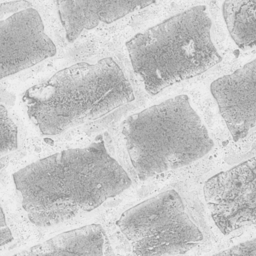} & \hspace{-4.0mm} \includegraphics[align=c, width=0.1\linewidth]{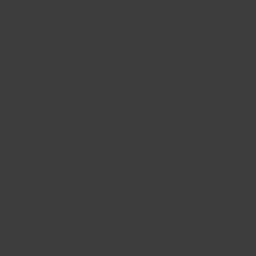} & \hspace{-4mm} \begin{sideways} \hspace{-3mm} \tiny{Specular} \end{sideways} \vspace{1mm} \\
\end{tabular} & \hspace{-4.0mm} \includegraphics[align=c, width=0.21\linewidth]{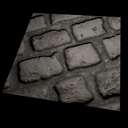} \vspace{1mm} \\
\begin{sideways}\hspace{-2.0mm}Ours\end{sideways} & \hspace{-4.0mm} \includegraphics[align=c, width=0.21\linewidth]{Figures/ComparisonAdaINPatchMatch/Inputs/558_brick_uneven_stones.png} & \begin{tabular} {cc}
\hspace{-4mm} \begin{sideways} \hspace{-3mm} \tiny{Normals} \end{sideways} & \hspace{-4.0mm} \includegraphics[align=c, width=0.045\linewidth]{Figures/ComparisonAdaINPatchMatch/Inputs/558_brick_uneven_stones_example_0.png} \vspace{1mm} \\
\hspace{-4mm} \begin{sideways} \hspace{-3mm} \tiny{Diffuse} \end{sideways} & \hspace{-4.0mm} \includegraphics[align=c, width=0.045\linewidth]{Figures/ComparisonAdaINPatchMatch/Inputs/558_brick_uneven_stones_example_1.png} \vspace{1mm} \\
\hspace{-4mm} \begin{sideways} \hspace{-3mm} \tiny{Rough} \end{sideways} & \hspace{-4.0mm} \includegraphics[align=c, width=0.045\linewidth]{Figures/ComparisonAdaINPatchMatch/Inputs/558_brick_uneven_stones_example_2.png} \vspace{1mm} \\
\hspace{-4mm} \begin{sideways} \hspace{-3mm} \tiny{Specu} \end{sideways} & \hspace{-4.0mm} \includegraphics[align=c, width=0.045\linewidth]{Figures/ComparisonAdaINPatchMatch/Inputs/558_brick_uneven_stones_example_3.png} \vspace{1mm} \\
\end{tabular} & \begin{tabular} {cccc}
\hspace{-4mm} \begin{sideways} \hspace{-3mm} \tiny{Normals} \end{sideways} & \hspace{-4.0mm} \includegraphics[align=c, width=0.1\linewidth]{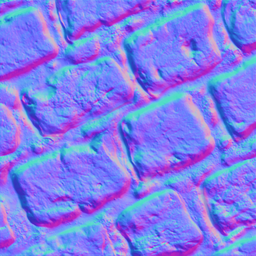} & \hspace{-4.0mm} \includegraphics[align=c, width=0.1\linewidth]{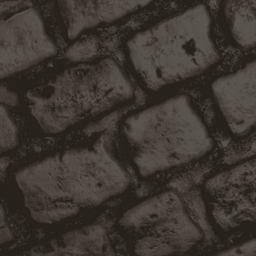} & \hspace{-4mm} \begin{sideways} \hspace{-3mm} \tiny{Diffuse} \end{sideways} \vspace{1mm} \\
\hspace{-4mm} \begin{sideways} \hspace{-3mm} \tiny{Roughness} \end{sideways} & \hspace{-4.0mm} \includegraphics[align=c, width=0.1\linewidth]{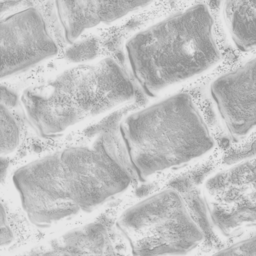} & \hspace{-4.0mm} \includegraphics[align=c, width=0.1\linewidth]{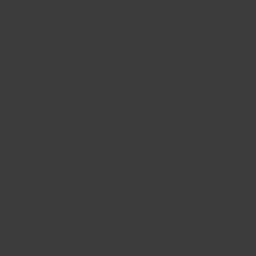} & \hspace{-4mm} \begin{sideways} \hspace{-3mm} \tiny{Specular} \end{sideways} \vspace{1mm} \\
\end{tabular} & \hspace{-4.0mm} \includegraphics[align=c, width=0.21\linewidth]{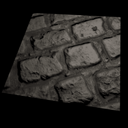} \vspace{1mm} \\
\end{tabular}
\caption{Comparison to neural style transfer \cite{huang17} and patch-based texture synthesis \cite{Fiser16-SIG}. Our method better transfers details of the surface compared to prior work, which either only captures global statistics (1st row) or struggles to generalize from a limited exemplar (2nd row).}
\label{fig:comparisonAdaINPatchMatch}
\end{figure*}

\paragraph*{Qualitative comparisons.}
Our approach is related to the method by Melendez et al.~\shortcite{Melendez12}, which transfers diffuse albedo and displacement maps using patch-based texture synthesis akin to image analogies \cite{Hertzmann2001Image}. We reproduced this approach with the state-of-the-art patch-based synthesis algorithm of Fi\v{s}er et al. \cite{Fiser16-SIG}, using the rendered SVBRDF as guidance. Note that since this algorithm was originally developed for style transfer, it assumes that the image to be synthesized only contains three color channels; to cope with this we ran their code on each SVBRDF parameter map separately. 
Fig.~\ref{fig:comparisonAdaINPatchMatch} shows results of this experiment; Patch-based synthesis lacks variety in the maps due to the limited information contained in a single exemplar. While more advanced synthesis algorithms exist to interpolate between limited exemplars \cite{TexSynth2015}, our deep learning solution natively generalizes the exemplar to the entire large-scale image. 

Fig.~\ref{fig:comparisonAdaINPatchMatch} also includes a comparison to AdaIN \cite{huang17}, a stylization algorithm based on deep learning that transfers statistics of deep features between an exemplar image and a target. Similarly to the above experiment, we applied the original implementation of the method on each SVBRDF parameter map separately. While this generic style transfer algorithm reproduces the overall color distribution of the maps, it misses many of the fine details. 

\begin{figure*}
\begin{tabular} {ccccc}
Method & HD input picture & Examplar & Results & Rendering \vspace{1mm} \\

\begin{sideways}\hspace{-6mm}\cite{Li17}\end{sideways} & \hspace{-4.0mm} \includegraphics[align=c, width=0.21\linewidth]{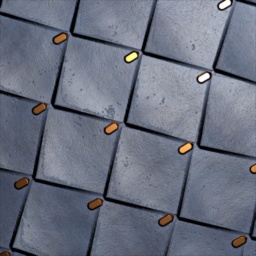} &  & \begin{tabular} {cccc}
\hspace{-4mm} \begin{sideways} \hspace{-3mm} \tiny{Normals} \end{sideways} & \hspace{-4.0mm} \includegraphics[align=c, width=0.1\linewidth]{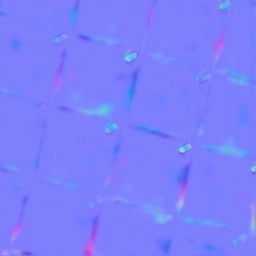} & \hspace{-4.0mm} \includegraphics[align=c, width=0.1\linewidth]{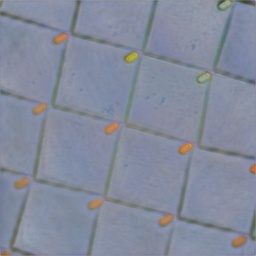} & \hspace{-4mm} \begin{sideways} \hspace{-3mm} \tiny{Diffuse} \end{sideways} \vspace{1mm} \\
\hspace{-4mm} \begin{sideways} \hspace{-3mm} \tiny{Roughness} \end{sideways} & \hspace{-4.0mm} \includegraphics[align=c, width=0.1\linewidth]{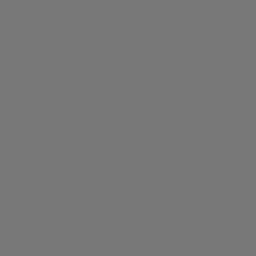} & \hspace{-4.0mm} \includegraphics[align=c, width=0.1\linewidth]{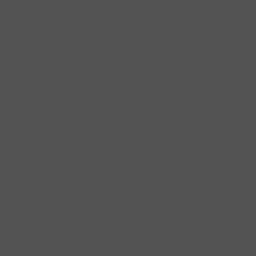} & \hspace{-4mm} \begin{sideways} \hspace{-3mm} \tiny{Specular} \end{sideways} \vspace{1mm} \\
\end{tabular} & \hspace{-4.0mm} \includegraphics[align=c, width=0.21\linewidth]{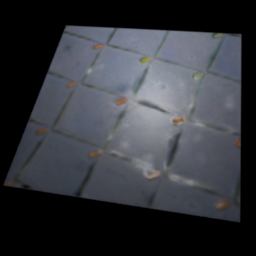} \vspace{1mm} \\

\begin{sideways}\hspace{-14mm}\cite{Deschaintre18} No flash\end{sideways} & \hspace{-4.0mm} \includegraphics[align=c, width=0.21\linewidth]{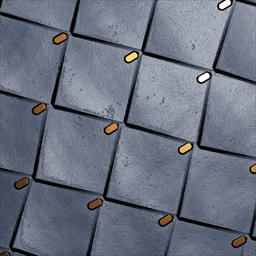} &  & \begin{tabular} {cccc}
\hspace{-4mm} \begin{sideways} \hspace{-3mm} \tiny{Normals} \end{sideways} & \hspace{-4.0mm} \includegraphics[align=c, width=0.1\linewidth]{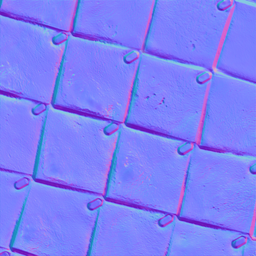} & \hspace{-4.0mm} \includegraphics[align=c, width=0.1\linewidth]{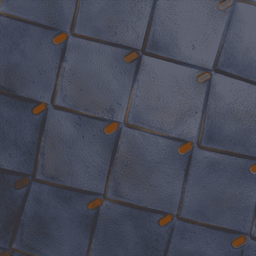} & \hspace{-4mm} \begin{sideways} \hspace{-3mm} \tiny{Diffuse} \end{sideways} \vspace{1mm} \\
\hspace{-4mm} \begin{sideways} \hspace{-3mm} \tiny{Roughness} \end{sideways} & \hspace{-4.0mm} \includegraphics[align=c, width=0.1\linewidth]{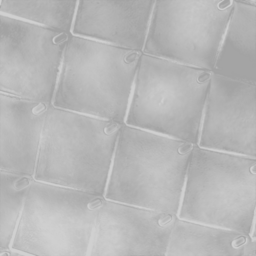} & \hspace{-4.0mm} \includegraphics[align=c, width=0.1\linewidth]{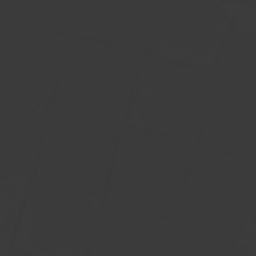} & \hspace{-4mm} \begin{sideways} \hspace{-3mm} \tiny{Specular} \end{sideways} \vspace{1mm} \\
\end{tabular} & \hspace{-4.0mm} \includegraphics[align=c, width=0.21\linewidth]{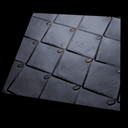} \vspace{1mm} \\

\begin{sideways}GT\end{sideways} &  &  & \begin{tabular} {cccc}
\hspace{-4mm} \begin{sideways} \hspace{-3mm} \tiny{Normals} \end{sideways} & \hspace{-4.0mm} \includegraphics[align=c, width=0.1\linewidth]{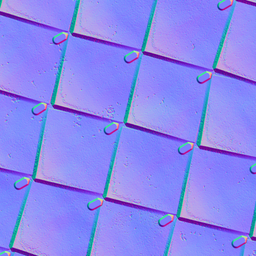} & \hspace{-4.0mm} \includegraphics[align=c, width=0.1\linewidth]{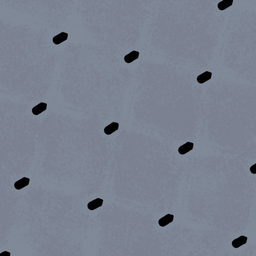} & \hspace{-4mm} \begin{sideways} \hspace{-3mm} \tiny{Diffuse} \end{sideways} \vspace{1mm} \\
\hspace{-4mm} \begin{sideways} \hspace{-3mm} \tiny{Roughness} \end{sideways} & \hspace{-4.0mm} \includegraphics[align=c, width=0.1\linewidth]{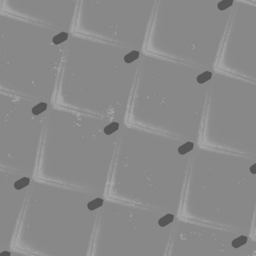} & \hspace{-4.0mm} \includegraphics[align=c, width=0.1\linewidth]{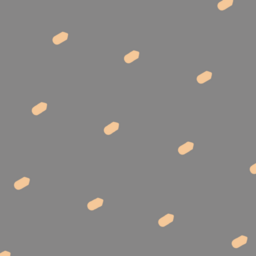} & \hspace{-4mm} \begin{sideways} \hspace{-3mm} \tiny{Specular} \end{sideways} \vspace{1mm} \\
\end{tabular} & \hspace{-4.0mm} \includegraphics[align=c, width=0.21\linewidth]{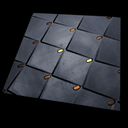} \vspace{1mm} \\

\begin{sideways}\hspace{-17mm}Ours, \cite{Deschaintre18} exemplars\end{sideways} & \hspace{-4.0mm} \includegraphics[align=c, width=0.21\linewidth]{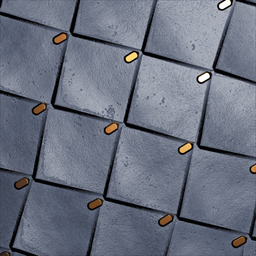} & \begin{tabular} {cc}
\hspace{-4mm} \begin{sideways} \hspace{-3mm} \tiny{Normals} \end{sideways} & \hspace{-4.0mm} \includegraphics[align=c, width=0.045\linewidth]{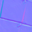} \vspace{1mm} \\
\hspace{-4mm} \begin{sideways} \hspace{-3mm} \tiny{Diffuse} \end{sideways} & \hspace{-4.0mm} \includegraphics[align=c, width=0.045\linewidth]{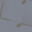} \vspace{1mm} \\
\hspace{-4mm} \begin{sideways} \hspace{-4mm} \tiny{Rough} \end{sideways} & \hspace{-4.0mm} \includegraphics[align=c, width=0.045\linewidth]{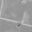} \vspace{1mm} \\
\hspace{-4mm} \begin{sideways} \hspace{-3mm} \tiny{Specular} \end{sideways} & \hspace{-4.0mm} \includegraphics[align=c, width=0.045\linewidth]{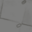} \vspace{1mm} \\
\end{tabular} & \begin{tabular} {cccc}
\hspace{-4mm} \begin{sideways} \hspace{-3mm} \tiny{Normals} \end{sideways} & \hspace{-4.0mm} \includegraphics[align=c, width=0.1\linewidth]{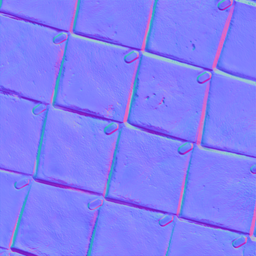} & \hspace{-4.0mm} \includegraphics[align=c, width=0.1\linewidth]{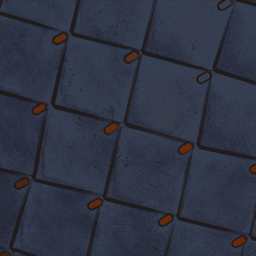} & \hspace{-4mm} \begin{sideways} \hspace{-3mm} \tiny{Diffuse} \end{sideways} \vspace{1mm} \\
\hspace{-4mm} \begin{sideways} \hspace{-3mm} \tiny{Roughness} \end{sideways} & \hspace{-4.0mm} \includegraphics[align=c, width=0.1\linewidth]{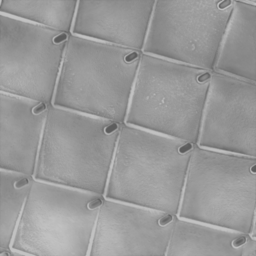} & \hspace{-4.0mm} \includegraphics[align=c, width=0.1\linewidth]{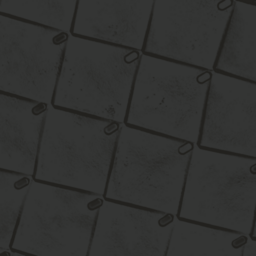} & \hspace{-4mm} \begin{sideways} \hspace{-3mm} \tiny{Specular} \end{sideways} \vspace{1mm} \\
\end{tabular} & \hspace{-4.0mm} \includegraphics[align=c, width=0.21\linewidth]{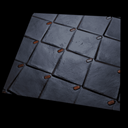} \vspace{1mm} \\

\end{tabular}
\caption{Comparison to the single-image methods of \protect\cite{Li17} and \protect \cite{Deschaintre18}. Thanks to a small exemplar, our method recovers more pronounced normal maps than the one by \protect\cite{Li17}, and also better captures the roughness of the small shiny metal plates, even though their specular strength remains underestimated. Also, since our method can process high-resolution images, it recovers finer details in the maps. Note that Li et al. use a different BRDF model than ours, so the values of their predicted maps shouldn't be directly compared to the ground truth maps.}
\label{fig:comparisonLiDeschaintre}
\end{figure*}

Finally, we provide in Fig.~\ref{fig:comparisonLiDeschaintre} a visual comparison to the recent deep learning methods for single-image SVBRDF capture by Li et al. \shortcite{Li17} and Deschaintre et al.~\shortcite{Deschaintre18}. 
While the method of Li et al. takes as input images captured under environment lighting, the original method of Deschaintre et al. assumes flash lighting, which is not compatible with the large-scale application scenarios we target. We thus re-trained their network on our training data to illustrate their performance on large-scale images taken without flash. Finally, for our method, we used the original method of Deschaintre et al. to recover SVBRDF exemplars from crops of the surface rendered under flash lighting, which emulates our on-site capture scenario.
Both prior methods struggle to recover the shininess of the little metallic plates. Our method better recovers these small shiny parts thanks to the provided exemplar.
In addition, our method can process large-scale images at high resolution, resulting in finer details in the SVBRDF maps.


\paragraph*{Quantitative comparisons.}
Table~\ref{tab:rmse} shows numerical comparisons to the single-image method of Deschaintre et al.~\shortcite{Deschaintre18} trained and tested on our data, and to the method by Li et al. \shortcite{Li17} applied on images rendered under environment lighting.
As in Fig.~\ref{fig:comparisonLiDeschaintre}, we obtained exemplars for our method by providing crops of the ground truth SVBRDF rendered under flash lighting to the original method of Deschaintre et al. In this setup, a single exemplar is enough to outperform competitors. In addition, we also provide the performance of our method when guided by ground truth exemplars, which can be seen as an upper-bound on the quality it can achieve.

Finally, Table~\ref{tab:rmse} (4th column) provides a numerical comparison to a version of our method inspired by the recent few-shot learning strategy proposed by Liu et al.~\shortcite{Liu_2019_ICCV}, who build on AdaIN to transfer style from multiple exemplars provided at test time. 
We adapted their approach to our context by processing each SVBRDF exemplar with
the encoder of Gao et al.~\shortcite{gao19} and by aggregating the resulting low-dimensional latent codes into a single code via max pooling. We next process this code with three fully-connected layers to produce parameters for several AdaIN layers that we use to transform the feature maps of the SVBRDF prediction network. The numerical evaluation reveals that the addition of AdaIN layers controled by the exemplars slightly improves performance over the baseline network of Deschaintre et al.~\shortcite{Deschaintre18} for some of the maps, but is largely inferior to our results obtained after fine-tuning this baseline on augmented exemplars.

\subsection{Limitations}
As with previous deep-learning based methods for material capture~\cite{Deschaintre18,li18}, 
we cannot handle cast shadows, or any other phenomenon that requires more than a normal/bump map.
Extending our approach to handle such cases, \emph{e.g.}, using a displacement map, would require
a much more complex differentiable renderer to handle 3D during training.
Similarly, our SVBRDF model and renderer are not designed to handle non-local effects like sub-surface scattering.

\begin{figure*}
\begin{tabular} {cccc}
HD input picture & Exemplar & Results & Rendering \vspace{1mm} \\
\hspace{-4.0mm} \includegraphics[align=c, width=0.22\linewidth]{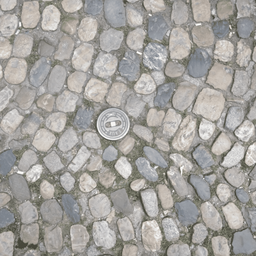} & \begin{tabular} {ccc}
\hspace{-4mm} \begin{sideways} \hspace{-3mm} \tiny{Normals} \end{sideways} & \hspace{-4.0mm} \includegraphics[align=c, width=0.05\linewidth]{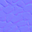} & \hspace{-4.0mm} \includegraphics[align=c, width=0.05\linewidth]{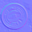} \vspace{1mm} \\
\hspace{-4mm} \begin{sideways} \hspace{-3mm} \tiny{Diffuse} \end{sideways} & \hspace{-4.0mm} \includegraphics[align=c, width=0.05\linewidth]{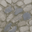} & \hspace{-4.0mm} \includegraphics[align=c, width=0.05\linewidth]{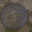} \vspace{1mm} \\
\hspace{-4mm} \begin{sideways} \hspace{-3mm} \tiny{Rough} \end{sideways} & \hspace{-4.0mm} \includegraphics[align=c, width=0.05\linewidth]{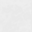} & \hspace{-4.0mm} \includegraphics[align=c, width=0.05\linewidth]{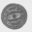} \vspace{1mm} \\
\hspace{-4mm} \begin{sideways} \hspace{-3mm} \tiny{Specular} \end{sideways} & \hspace{-4.0mm} \includegraphics[align=c, width=0.05\linewidth]{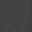} & \hspace{-4.0mm} \includegraphics[align=c, width=0.05\linewidth]{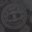} \vspace{1mm} \\
\end{tabular} & \begin{tabular} {cccc}
\hspace{-4mm} \begin{sideways} \hspace{-3mm} \tiny{Normals} \end{sideways} & \hspace{-4.0mm} \includegraphics[align=c, width=0.11\linewidth]{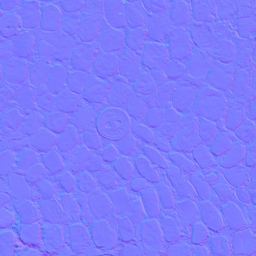} & \hspace{-4.0mm} \includegraphics[align=c, width=0.11\linewidth]{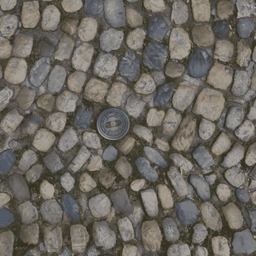} & \hspace{-4mm} \begin{sideways} \hspace{-3mm} \tiny{Diffuse} \end{sideways} \vspace{1mm} \\
\hspace{-4mm} \begin{sideways} \hspace{-3mm} \tiny{Roughness} \end{sideways} & \hspace{-4.0mm} \includegraphics[align=c, width=0.11\linewidth]{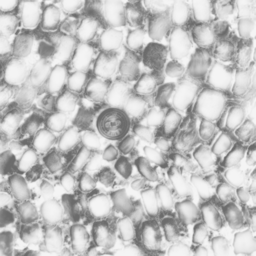} & \hspace{-4.0mm} \includegraphics[align=c, width=0.11\linewidth]{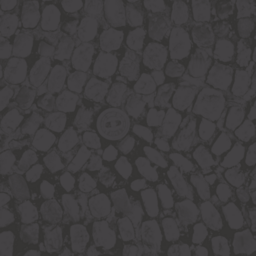} & \hspace{-4mm} \begin{sideways} \hspace{-3mm} \tiny{Specular} \end{sideways} \vspace{1mm} \\
\end{tabular} & \hspace{-4.0mm} \includegraphics[align=c, width=0.22\linewidth]{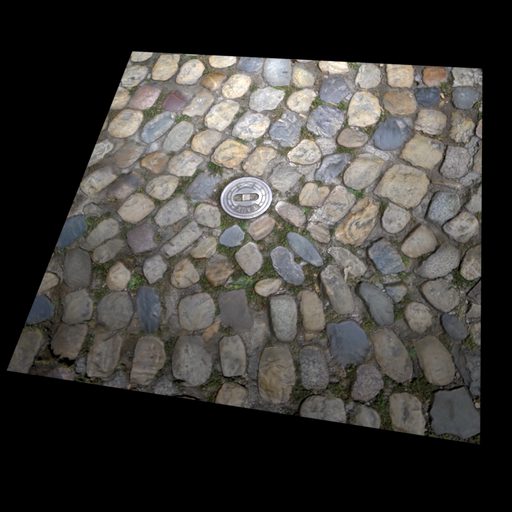} \vspace{1mm} \\
\end{tabular}
\vspace{-2mm}
\caption{Limitation. Our method can have difficulty distinguishing materials with similar colors and texture, such as this shiny metal disk that has a similar appearance to some of the dark rough stones.}
\label{fig:limitations1}
\end{figure*}

\begin{figure*}
\begin{tabular} {cccc}
HD input picture & Exemplar & Results & Rendering \vspace{1mm} \\
\hspace{-4.0mm} \includegraphics[align=c, width=0.44\linewidth]{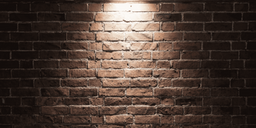} & \begin{tabular} {cc}
\hspace{-4mm} \begin{sideways} \hspace{-3mm} \tiny{Diffuse} \end{sideways} & \hspace{-4.0mm} \includegraphics[align=c, width=0.07\linewidth]{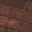} \vspace{1mm} \\
\hspace{-4mm} \begin{sideways} \hspace{-3mm} \tiny{Rough} \end{sideways} & \hspace{-4.0mm} \includegraphics[align=c, width=0.07\linewidth]{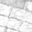} \vspace{1mm} \\
\hspace{-4mm} \begin{sideways} \hspace{-3mm} \tiny{Specular} \end{sideways} & \hspace{-4.0mm} \includegraphics[align=c, width=0.07\linewidth]{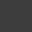} \vspace{1mm} \\
\end{tabular} & \begin{tabular} {cc}
\hspace{-4mm} \begin{sideways} \hspace{-3mm} \tiny{Diffuse} \end{sideways} & \hspace{-4.0mm} \includegraphics[align=c, width=0.14\linewidth]{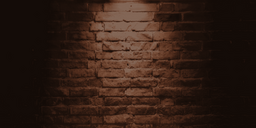} \vspace{1mm} \\
\hspace{-4mm} \begin{sideways} \hspace{-3mm} \tiny{Roughness} \end{sideways} & \hspace{-4.0mm} \includegraphics[align=c, width=0.14\linewidth]{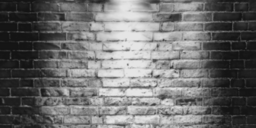} \vspace{1mm} \\
\hspace{-4mm} \begin{sideways} \hspace{-3mm} \tiny{Specular} \end{sideways} & \hspace{-4.0mm} \includegraphics[align=c, width=0.14\linewidth]{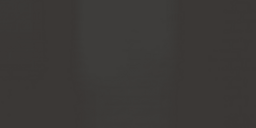} \vspace{1mm} \\
\end{tabular} & \hspace{-4.0mm} \includegraphics[align=c, width=0.22\linewidth]{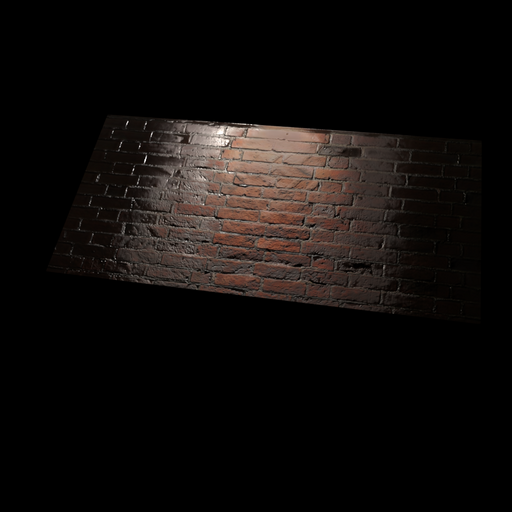} \vspace{1mm} \\
\end{tabular}
\vspace{-2mm}
\caption{Limitations. Our method is not designed to handle large illumination gradients over the surface.}
\label{fig:limitations2}
\end{figure*}

Despite the strong ability of deep learning to extract discriminative features, our method sometimes has difficulty distinguishing different materials that share similar colors and textures. This is the case in Fig.~\ref{fig:limitations1}, where the shininess of the small metal disk is transferred to some of the stones that have a similar appearance in the input picture.
Our method also assumes that the large-scale input is captured under largely uniform lighting.
When this is not the case, large illumination gradients pollute the SVBRDF maps, as shown in Fig.~\ref{fig:limitations2}. \NEW{Nevertheless, our method is robust to localized highlights, as some occur in the training set (see synthetic materials in supplemental materials for typical examples).}

\NEW{Finally, while there is a theoretical limitation to the scale difference that our method can handle between the exemplar and large-scale input to correctly transfer the materials, we never encountered this problem in our tests.}



\section{Conclusion}
Our method alleviates inherent limitations of flash-based material acquisition methods, namely limited scale, low resolution, and lack of user control. By complementing the input image with one or a few exemplars, our approach can recover SVBRDFs of much larger surfaces, at high resolution and arbitrary aspect ratio. Furthermore, our method greatly increases the creative freedom of material designers by letting them create plausible SVBRDFs from existing photographs with high-level control on their constituent materials. We achieved all these benefits thanks to a surprisingly simple fine-tuning strategy, which we believe to be directly applicable to other capture and design tasks based on deep learning. 



\section*{Acknowledgments}
We thank Simon Rodriguez for his help with video editing. This work was partially funded by an ANRT(\url{http://www.anrt.asso.fr/en}) CIFRE scholarship between Inria and Optis for Ansys, ERC Advanced Grant FUNGRAPH (No. 788065, \url{http://fungraph.inria.fr}), EPSRC Early Career Fellowship (EP/N006259/1) and by software donations from Adobe. The authors are grateful to Inria Sophia Antipolis - Méditerranée "Nef" computation cluster for providing resources and support (\url{https://wiki.inria.fr/ClustersSophia/Clusters_Home}). 

\bibliographystyle{eg-alpha-doi}
\bibliography{bibliography/bibliography} 
\end{document}